\documentclass[journal,twoside,web]{ieeecolor}
\usepackage{generic}
\usepackage{cite}
\usepackage{amsmath,amssymb,amsfonts}

\newtheorem{definition}{Definition}
\newtheorem{theorem}{Theorem}
\newtheorem{lemma}{Lemma}

\newtheorem{proposition}{Proposition}
\newtheorem{problem}{Problem}

\usepackage{graphicx}
\usepackage{textcomp}
\usepackage{hyperref}
\usepackage[noend,ruled,linesnumbered]{algorithm2e}

\def\BibTeX{{\rm B\kern-.05em{\sc i\kern-.025em b}\kern-.08em
    T\kern-.1667em\lower.7ex\hbox{E}\kern-.125emX}}

\makeatletter
\def\custom@bluehead{%
  \vbox{%
    \hbox to \textwidth{%
  \normalfont\scriptsize\sffamily\color{black}%
  \ifodd\c@page
    \hfil\thepage%
  \else
    \thepage\hfil%
  \fi
}%
    \kern2pt
    {\color{subsectioncolor}%
     \hrule width\textwidth height1pt}%
  }%
}

\g@addto@macro\ps@titlepagestyle{%
  \let\@oddhead\custom@bluehead
  \let\@evenhead\custom@bluehead
}

\g@addto@macro\ps@headings{%
  \let\@oddhead\custom@bluehead
  \let\@evenhead\custom@bluehead
}

\makeatother

\begin{document}

\title{A Scalable Decentralized Event-Triggered Topology-Switching Protocol for Verifiable Safe Control in Multi-Agent Systems}

\author{Amber~He~Wei,~\IEEEmembership{Student Member, IEEE,}
Shiyu~Cheng,~\IEEEmembership{Graduate Student Member, IEEE,}
Andrew~Clark,~\IEEEmembership{Senior Member, IEEE,}
and~Bruno~Sinopoli,~\IEEEmembership{Fellow, IEEE}
\thanks{A. H. Wei, S. Cheng and A. Clark are with the Department of Electrical and Systems Engineering, Washington University in St. Louis, St. Louis, MO, USA. Email: \{hwei, cheng.shiyu, andrewclark\}@wustl.edu}
\thanks{B. Sinopoli is with the School of Electrical, Computer and Energy Engineering, Arizona State University, Tempe, AZ, USA. Email: bruno.sinopoli@asu.edu}
}

\maketitle

\begin{abstract}
This paper develops a decentralized event-triggered topology-switching protocol that allows agents to disconnect from the communication network while preserving state safety and excluding Zeno behavior. We establish a robust barrier function condition that guarantees safety as long as the estimation error remains within a prescribed threshold. Each agent uses locally computable certificates to determine when to connect and disconnect from the network. The certificate construction improves scalability by reducing the complexity of topology-dependent computations from $O(2^N)$ to $O(N)$, where $N$ is the number of agents. We formally prove that the protocol ensures the estimation error does not exceed the prescribed threshold and is non-Zeno. An alternating method computes an error-weighting matrix and a quadratic barrier function that guarantee safety under the proposed protocol. Finally, we evaluate the protocol on four interconnected microgrids. Compared with maintaining network connectivity at all times, the protocol reduces the time agents remain connected while keeping phase and frequency within their safe operating region.
\end{abstract}

\begin{IEEEkeywords}
Multi-agent systems, networked control, control barrier function, event-triggered control.
\end{IEEEkeywords}

%%%%%%%%%%%%%%%%%%%%%%%%%%%%%%%%%%%%%%%%%%%%%%%%%%%%%%%%%%%%%%%%%%%%%%%%%%%%%%%%
\section{Introduction}
\label{sec:introduction}

Over the past few decades, decentralized control of multi-agent systems has received increasing attention in applications such as automated vehicles~\cite{sabouni2023optimal}, water distribution networks~\cite{johansson2000}, mobile robots~\cite{Wang_2016,luo2020multi}, and power grids~\cite{9507261}. In these systems, agents coordinate their actions to achieve a global objective through local communication and feedback control~\cite{bullo2009distributed}. Many of these applications are safety-critical because failures can cause serious damage to physical infrastructure and endanger human lives. Control actions must therefore satisfy safety constraints during operation. These constraints can depend on the states of multiple agents. For example, maintaining a minimum distance between two robots requires information about both robots' positions~\cite{an2023event}.

In a decentralized setting, each agent measures only part of the global state and uses an estimate of the global state to compute its control input. If an agent's own measurements are insufficient, the agent's state estimate may drift away from the true global state over time, increasing its estimation error. As the estimation error grows, control actions based on the state estimate may cause the system to violate safety constraints. To prevent such violations, a common approach is to derive a safety condition that holds for every estimation error within a prescribed threshold~\cite{9636584,9867262}. If the estimation error remains within this threshold, the condition guarantees state safety. When an agent cannot keep its estimation error within this threshold using only its own measurements, it must communicate with other agents to obtain information that improves its state estimate.

To exchange information, agents form a communication network. When an agent connects to the network, it can receive new information from other agents and broadcast its own information to them. Maintaining the connections consumes bandwidth and energy~\cite{rajaraman2002topology,peng2013event} and increases the risk of cyberattacks~\cite{nowzari2019event}. These costs and risks motivate reducing communication. Event-triggered control (ETC) is widely used to reduce communication overhead in multi-agent systems by exchanging information or updating control inputs only when prescribed triggering conditions are met~\cite{6425820,nowzari2019event,ge2021dynamic,deng2022distributed}. Traditional ETC requires agents to remain connected to the network at all times so that the state of each agent can be accessed at any time to evaluate the triggering condition. Once the current state violates the triggering condition, the protocol activates communication to improve the agent's state estimate. In this case, although communication is reduced, agents still maintain network connections at all times, which incurs network management overhead and creates an attack surface for cyber adversaries.

Prior work~\cite{10360262} addressed this drawback by proposing a decentralized ETC protocol that allows agents to disconnect from the network when communication is not needed, while maintaining system stability. Each agent independently determines when to reconnect by evaluating a prescribed triggering condition. While disconnected, an agent updates its estimate using only its own measurements. The protocol guarantees convergence of estimation errors to a prescribed region but allows them to leave this region temporarily before converging back to it after reconnection. This convergence guarantee does not ensure state safety because the estimation error may exceed the prescribed threshold before convergence.

In this work, we address the gap in guaranteeing state safety under intermittent connectivity by developing a decentralized event-triggered topology-switching protocol. Here, we use forward invariance to guarantee safety: a state trajectory that starts in a safe set remains in that set for all time. Control barrier functions (CBFs) are widely used to certify state safety through forward invariance of a safe set~\cite{ames2017control,8796030,Wang_2016,sabouni2023optimal,wu2016safety}. We use a barrier function to certify forward invariance of a subset of the prescribed safe set under the given controller when the estimation error remains within a prescribed threshold. Under our protocol, each agent computes an upper bound on the estimation error from its available topology information. By comparing its estimation-error bound with the prescribed threshold, each agent independently determines when to connect and disconnect.

Our main contributions are as follows:
\begin{itemize}
    \item We establish a sufficient condition for state safety that combines a barrier condition with the requirement that the estimation error remain within a prescribed threshold.
    \item We prove that the protocol is non-Zeno and keeps the estimation error within the prescribed threshold at all times, thereby guaranteeing state safety.
    \item We adopt an alternating method to compute an error-weighting matrix that defines the prescribed estimation-error threshold and a barrier function that certifies state safety.
    \item We evaluate the protocol on a nonlinear four-microgrid system, demonstrating that phase and frequency deviations remain within prescribed limits under intermittent connectivity.
\end{itemize}

Compared with our conference paper~\cite{wei2026verifiablysafedecentralizedcontrol}, this work develops a more scalable topology-switching protocol. Using each agent's available topology information, the protocol reduces the number of topology-dependent calculations from $O(2^N)$ to $O(N)$, where $N$ is the number of agents. We also prove that the protocol is non-Zeno and adopt an alternating optimization method to compute an error-weighting matrix and a barrier function satisfying the required conditions, rather than assuming these quantities are given.

The paper is organized as follows. Section~II presents the system model and preliminaries. Section~III formulates the safety requirement and derives the robust barrier-certificate conditions. Section~IV develops the topology-switching protocol and proves that it preserves safety and excludes Zeno behavior. Section~V presents the alternating method. Section~VI presents simulation, and Section~VII concludes the paper.
\section{Model and Preliminaries}
\label{sec:model}

\subsection{Notation}

Let $\mathbb R_{\geq0}$ and $\mathbb R_{>0}$ denote the sets of nonnegative and positive real numbers. Let $\mathbb S^n\triangleq\{X\in\mathbb R^{n\times n}:X=X^\top\}$ denote the set of real symmetric matrices. For $X,Y\in\mathbb S^n$, write $X\succeq Y$ if $X-Y$ is positive semidefinite and $X\succ Y$ if $X-Y$ is positive definite. Accordingly, define
$\mathbb S_+^n\triangleq\{X\in\mathbb S^n:X\succeq0\}$ and
$\mathbb S_{++}^n\triangleq\{X\in\mathbb S^n:X\succ0\}$.
For $X\in\mathbb S^n$, let $\lambda_{\max}(X)$ denote its largest eigenvalue. For $X\in\mathbb S^n$ with $n\geq2$, let $\lambda_2(X)$ denote its second-smallest eigenvalue. For any square matrix $A$, let $\operatorname{tr}(A)$ denote its trace. Let $I_n$ denote the $n\times n$ identity matrix and $0_{p\times q}$ the $p\times q$ zero matrix. For matrices $A_1,\ldots,A_m$, $\operatorname{diag}(A_1,\ldots,A_m)$ is the block-diagonal matrix whose diagonal blocks, in order, are $A_1,\ldots,A_m$. The symbol $\otimes$ denotes the Kronecker product.

For vectors, $\|\cdot\|_2$ is the Euclidean norm. For matrices, $\|\cdot\|_2$ is the induced $2$-norm. The symbol $|\cdot|$ denotes absolute value for scalars and cardinality for finite sets. For a set $\mathcal S$, $\partial\mathcal S$ denotes its boundary. 

For a real-valued function $g$ defined in a neighborhood of $t$, its upper-right Dini derivative is $D^+g(t)\triangleq\limsup_{h\to0^+}\frac{g(t+h)-g(t)}{h}.$ When the limits exist, the left and right limits of $g$ at $t$ are
$g(t^-)\triangleq\lim_{h\to0^+}g(t-h),
~
g(t^+)\triangleq\lim_{h\to0^+}g(t+h).$

\subsection{System Model}

We consider $N$ agents, with each agent indexed by
$i\in\mathcal N\triangleq\{1,\ldots,N\}$. The global dynamics follow the
continuous-time linear time-invariant (LTI) model
\begin{equation}\label{system}
\begin{aligned}
\dot x(t)&=Ax(t)+Bu(t)+w(t),\\
y(t)&=Cx(t)+v(t).
\end{aligned}
\end{equation}
The variables in \eqref{system} are defined as follows:
\begin{itemize}
    \item $x(t)\triangleq[x_1(t)^\top,\ldots,x_N(t)^\top]^\top\in\mathbb R^n$ is the global state, where $x_i(t)\in\mathbb R^{n_i}$ and $\sum_{i=1}^N n_i=n$.

    \item $u(t)\triangleq[u_1(t)^\top,\ldots,u_N(t)^\top]^\top\in\mathbb R^m$ is the control input,
    where $u_i(t)\in\mathbb R^{m_i}$ and $\sum_{i=1}^N m_i=m$.

    \item $y(t)\triangleq[y_1(t)^\top,\ldots,y_N(t)^\top]^\top\in\mathbb R^p$ is the measurement output,
    where $y_i(t)\in\mathbb R^{p_i}$ and $\sum_{i=1}^N p_i=p$.

    \item $w(t)\in\mathbb R^n$ is the process disturbance.
    $v(t)\triangleq[v_1(t)^\top,\ldots,v_N(t)^\top]^\top\in\mathbb R^p$ is the measurement noise,
    where $v_i(t)\in\mathbb R^{p_i}$.
\end{itemize}
The matrices $A\in\mathbb R^{n\times n}$,
$B\triangleq[B_1~\cdots~B_N]\in\mathbb R^{n\times m}$,
$C\triangleq[C_1^\top,\ldots,C_N^\top]^\top\in\mathbb R^{p\times n}$,
with $C_i\in\mathbb R^{p_i\times n}$ and $B_i\in\mathbb R^{n\times m_i}$, are known and constant.
Agent $i$ accesses its local measurement
$y_i(t)=C_i x(t)+v_i(t)$. 

We assume that $w(t)\in\mathcal W$, $v(t)\in\mathcal V$
for all $t\geq0$, where
\begin{equation}\label{eq:disturbance-set}
\mathcal W\triangleq
\{w\in\mathbb R^n:w^\top Q_w w\leq1\},
\end{equation}
\begin{equation}\label{eq:noise-set}
\mathcal V\triangleq
\{v\in\mathbb R^p:v^\top R_v v\leq1\},
\end{equation}
The matrices $Q_w\in\mathbb S_{++}^n$ and $R_v=\operatorname{diag}(R_{v,1},\ldots,R_{v,N})\in\mathbb S_{++}^p$ are given, where $R_{v,i}\in\mathbb S_{++}^{p_i}$ for each $i\in\mathcal N$.

\subsection{Communication Model}
\label{subsec:communication-model}
Let $\bar{\mathcal G}\triangleq(\mathcal N,\bar{\mathcal E})$ be a fixed, connected, undirected graph. The communication topology at time $t$ is modeled by a spanning subgraph $\mathcal G(t)\triangleq(\mathcal N,\mathcal E(t))$ of $\bar{\mathcal G}$, meaning that $\mathcal G(t)$ and $\bar{\mathcal G}$ have the same vertex set $\mathcal N$ and $\mathcal E(t)\subseteq\bar{\mathcal E}$ for all $t\geq0$.

Let $\mathcal A(t)=[a_{ij}(t)]\in\mathbb R^{N\times N}$ be the symmetric adjacency matrix of $\mathcal G(t)$. If $(i,j)\in\mathcal E(t)$, then $a_{ij}(t)=1$; if $(i,j)\notin\mathcal E(t)$, then $a_{ij}(t)=0$. The diagonal entries satisfy $a_{ii}(t)=0$. An edge $(i,j)$ is called an \emph{active communication edge} at time $t$ if $a_{ij}(t)=1$, indicating that direct communication between agents $i$ and $j$ is enabled. The \emph{active neighbors} of agent $i$ are the agents that share an active communication edge with $i$. Each agent $i$ has access to its active communication edges. Its \emph{active neighboring set} is defined as
\begin{equation}\label{eq:active-neighbors}
    \mathcal N_i(t)\triangleq
    \{j\in\mathcal N\setminus\{i\}:a_{ij}(t)=1\}.
\end{equation}

To reduce communication overhead, each agent determines when to connect or disconnect and operates in one of the following modes:
\begin{itemize}
    \item \textbf{Disconnected mode (DM):} Agent $i$ is unavailable for communication and updates its state estimate using only its local measurements $y_i(t)$.

    \item \textbf{Connected mode (CM):} Agent $i$ is available for communication. If $\mathcal N_i(t)\neq\varnothing$, it exchanges information with its active neighbors and updates its state estimate using its local measurements $y_i(t)$ and information received from its active neighbors. If $\mathcal N_i(t)=\varnothing$, it updates its state estimate using only $y_i(t)$ but remains available for communication. An edge $(i,j)\in\bar{\mathcal E}$ can be active only if both agents are in connected mode. Thus, an agent in connected mode with no active neighbors remains available for communication and can exchange information with agent $j$ when $j$ enters connected mode and $(i,j)$ becomes active.
\end{itemize}

For an agent $i$ in connected mode, its connected component consists of $i$ and all agents reachable from $i$ through paths of active communication edges. A path from $i$ to $j$ at time $t$ is a sequence of distinct agents $i=i_0,i_1,\ldots,i_k=j$ satisfying $a_{i_{r-1}i_r}(t)=1$ for every $r=1,\ldots,k$. We assume that agent $i$ knows the vertex set and edge set of its connected component.

Let $\mathcal L(t)=[\ell_{ij}(t)]\in\mathbb R^{N\times N}$ denote the graph Laplacian associated with $\mathcal G(t)$, where $\ell_{ii}(t)=|\mathcal N_i(t)|$ and $\ell_{ij}(t)=-a_{ij}(t)$ for $i\neq j$. Let $\bar{\mathcal L}\in\mathbb R^{N\times N}$ denote the Laplacian of $\bar{\mathcal G}$. Since $\bar{\mathcal G}$ is connected, $\bar{\mathcal L}$ is positive semidefinite with a simple eigenvalue at 0 and $\lambda_2(\bar{\mathcal L})>0$. Moreover, $\mathcal E(t)\subseteq\bar{\mathcal E}$ implies $0\preceq\mathcal L(t)\preceq\bar{\mathcal L}$ for all $t\geq0$.

\subsection{Decentralized State Estimation}

The controller and observer, together with the resulting error dynamics
in Section~\ref{subsec:error-dynamics}, follow~\cite{10360262}. 

Each agent $i$ applies the control law $u_i(t)=K_i\hat{x}_i(t)$,
where $K_i\in\mathbb R^{m_i\times n}$ is a given local gain matrix
and $\hat{x}_i(t)\in\mathbb R^n$ is its local estimate of the global
state. Define the stacked gain
$K\triangleq[K_1^\top,\ldots,K_N^\top]^\top\in\mathbb R^{m\times n}$
and the nominal closed-loop matrix
$A_{bk}\triangleq A+BK\in\mathbb R^{n\times n}$.

The observer gain
$L\triangleq[L_1~\cdots~L_N]\in\mathbb R^{n\times p}$, with
$L_i\in\mathbb R^{n\times p_i}$, is chosen such that $A-LC$ is Hurwitz.
Agent $i$ has the mode-dependent observer gain
$\bar L_i(\ell_{ii}(t))\in\mathbb R^{n\times p_i}$ as
\begin{equation}
\bar{L}_i(\ell_{ii}(t))
\triangleq
\begin{cases}
\hat L_i, & \ell_{ii}(t)=0,\\
N L_i, & \ell_{ii}(t)>0,
\end{cases}
\label{mode-dependent gain}
\end{equation}
where $\hat L_i\triangleq T_i[L_i^o,0]^\top
\in\mathbb R^{n\times p_i}$. To construct $\hat L_i$,
the observability decomposition of $(A,C_i)$ gives an invertible transformation matrix
$T_i\in\mathbb R^{n\times n}$ such that
\begin{equation*}
\begin{aligned}
T_i^{-1} A T_i &=
\begin{bmatrix}
A_i^o & 0 \\
A_i^{21} & A_i^{\bar{o}}
\end{bmatrix},~
C_i T_i = \begin{bmatrix} C_i^o & 0 \end{bmatrix},
\end{aligned}
\end{equation*}
Let $n_i^o$ denote the dimension of the observable subspace of
$(A,C_i)$, with $n_i^o\leq n$. The matrices $A_i^o\in\mathbb R^{n_i^o\times n_i^o}$ and
$C_i^o\in\mathbb R^{p_i\times n_i^o}$ describe the observable
subsystem, $A_i^{\bar{o}}\in
\mathbb R^{(n-n_i^o)\times(n-n_i^o)}$ describes the
unobservable subsystem, and
$A_i^{21}\in\mathbb R^{(n-n_i^o)\times n_i^o}$ is the
off-diagonal block. The pair $(A_i^o,C_i^o)$ is observable, and
gain $L_i^o\in\mathbb R^{n_i^o\times p_i}$ is chosen such that
$A_i^o-L_i^o C_i^o$ is Hurwitz. 

We define \emph{the locally available information} of agent $i$ at time $t$ to consist of its local measurement $y_i(t)$, its own global-state estimate $\hat{x}_i(t)$, and its active-neighbor set $\mathcal N_i(t)$. When agent $i$ is in connected mode, this information also includes the estimates $\hat{x}_j(t)$ received from active neighbors $j\in\mathcal N_i(t)$ and the topology of its connected component. By this information, the decentralized observer dynamics for agent $i$ follows
\begin{equation}
\begin{aligned}
\dot{\hat{x}}_i(t) ={}& A_{bk}\hat{x}_i(t)
+ \bar{L}_i(\ell_{ii}(t))\bigl(y_i(t) - C_i\hat{x}_i(t)\bigr) \\
& + \eta \sum_{j=1}^N a_{ij}(t)
\bigl(\hat{x}_j(t) - \hat{x}_i(t)\bigr),
\end{aligned}
\end{equation}
where $\eta \ge 0$ is the network coupling gain, and $\ell_{ii}(t)$ is the
$i$-th diagonal entry of $\mathcal{L}(t)$.

\subsection{Error Dynamics and Closed-Loop Dynamics}
\label{subsec:error-dynamics}

We define agent $i$'s estimation error of the global state as
$e_i(t) \triangleq x(t) - \hat{x}_i(t) \in \mathbb{R}^n$. The error dynamics is given by
\begin{align*}
\dot{e}_i(t) ={}& \bigl(A_{bk} - \bar{L}_i(\ell_{ii}(t))C_i\bigr)e_i(t)
+ w(t) - \bar{L}_i(\ell_{ii}(t))v_i(t) \\
& - \sum_{j=1}^N \bigl(B_j K_j + \eta\,\ell_{ij}(t)\bigr)e_j(t).
\end{align*}
By stacking the agents' global-state estimation errors as
$e(t) \triangleq \bigl[e_1(t)^\top, \dots, e_N(t)^\top\bigr]^\top
\in\mathbb{R}^{Nn},$
we compactly write the stacked error dynamics as
\begin{equation}
\begin{aligned}
\label{eq:dynamics-error}
\dot{e}(t) ={}&
\underbrace{\bigl(F(\mathcal{L}(t))
- \eta(\mathcal{L}(t) \otimes I_n)\bigr)}_{A_e(\mathcal{L}(t))} e(t) \\
& + \mathcal{I} w(t) - J(\mathcal{L}(t)) v(t),
\end{aligned}
\end{equation}
where
$\mathcal{I} \triangleq
\bigl[I_n^\top, \dots, I_n^\top\bigr]^\top
\in\mathbb{R}^{Nn \times n},$
$J(\mathcal{L}(t)) \triangleq
\operatorname{diag}\bigl(
\bar L_1(\ell_{11}(t)), \dots, \bar L_N(\ell_{NN}(t))
\bigr),$
and $F(\mathcal{L}(t)) \in \mathbb{R}^{Nn \times Nn}$ is a block matrix with
block entries $F_{ij}(\mathcal{L}(t)) \in \mathbb{R}^{n \times n}$ as
\begin{equation}
F_{ij}(\mathcal{L}(t)) \triangleq
\scalebox{0.9}{$
\begin{cases}
A_{bk} - \bar{L}_i(\ell_{ii}(t))C_i - B_i K_i,
& \text{if } i = j,\\[4pt]
-B_j K_j, & \text{if } i \neq j.
\end{cases}$}
\label{Fij}
\end{equation}
Finally, substituting the local control inputs
$u_i(t) = K_i x(t) - K_i e_i(t)$ into the system model yields the global
closed-loop state dynamics
\begin{equation}\label{eq:closed_loop_dynamics}
\dot{x}(t) = A_{bk}x(t) - Ee(t) + w(t),
\end{equation}
where
$E \triangleq \begin{bmatrix} B_1 K_1 & \cdots & B_N K_N \end{bmatrix}
\in\mathbb{R}^{n \times Nn}$.

\subsection{Barrier Certificate}
\label{subsec:barrier-certificate}

Consider a dynamical system $\dot{x}(t)=f(x(t))$, where $f:\mathbb R^n\to\mathbb R^n$ is locally Lipschitz. Let $\phi:\mathbb R^n\to\mathbb R$ be continuously differentiable and define its zero-superlevel set as $\mathcal S\triangleq\{x\in\mathbb R^n:\phi(x)\geq0\}$. The derivative of $\phi$ along the vector field $f$ is given by $L_f\phi(x)\triangleq\nabla\phi(x)^\top f(x)$. An extended class-$\mathcal K$ function
$\alpha_\phi:(-a_\phi,b_\phi)\to\mathbb R$,
for some $a_\phi,b_\phi>0$, is continuous, strictly increasing,
and satisfies $\alpha_\phi(0)=0$. Assume that $\nabla\phi(x)\neq0$ for all
$x\in\partial\mathcal S$. The following definition follows~\cite{8796030}.

\begin{definition}
\label{def:barrier-certificate}
The function $\phi$ is a barrier certificate for $\mathcal S$ if
there exists an extended class-$\mathcal K$ function $\alpha_\phi$
such that
\begin{equation}
L_f\phi(x)+\alpha_\phi\bigl(\phi(x)\bigr)\geq0,
    \qquad \text{for all }x\in\mathcal S.
    \label{eq:general-bc-condition}
\end{equation}
\end{definition}
Let $x(t)$ be the unique solution of $\dot{x}(t)=f(x(t))$
with $x(0)\in\mathcal S$, and assume that it exists for all
$t\geq0$. If \eqref{eq:general-bc-condition} holds, then
$\mathcal S$ is forward invariant and $x(t)\in\mathcal S$
for all $t\geq0$.
\section{Problem Formulation}
\label{sec:formulation}

We first specify the safety requirement and develop sufficient
conditions for state safety under bounded estimation error. We then formulate the problem of developing a decentralized topology-switching protocol that ensures the estimation error does not exceed a prescribed threshold.

\subsection{Problem Setup}
\label{subsec:problem-statement}

The state is required to satisfy $x(t)\in\mathcal C$ for all $t\geq0$, where
\begin{equation}
    \mathcal C \triangleq \{x\in\mathbb R^n:h(x)\geq0\}
    \label{eq:safe-region}
\end{equation}
is a prescribed safe set and $h:\mathbb R^n\to\mathbb R$ is a given polynomial. The initial state is assumed to satisfy $x(0)\in\mathcal C$. However, the prescribed set $\mathcal C$ need not be forward invariant under \eqref{eq:closed_loop_dynamics}. We therefore seek a subset of $\mathcal C$ that contains the initial state and remains forward invariant as long as the estimation error does not exceed a prescribed threshold. To construct this subset, we introduce a polynomial barrier function $b:\mathbb R^n\to\mathbb R$ and define
\begin{equation}
    \mathcal B \triangleq \{x\in\mathbb R^n:b(x)\geq0\}
    \label{eq:barrier-set}
\end{equation}
such that $x(0)\in\mathcal B$ and $\mathcal B\subseteq\mathcal C$.

To quantify the estimation error, we define
\begin{equation}
    V_e(e)\triangleq e^\top\bar P e,
    \label{eq:error-certificate}
\end{equation}
where the global error-weighting matrix is
\begin{equation}
    \bar P\triangleq I_N\otimes P_e\in\mathbb S_{++}^{Nn},
    \label{eq:global-error-weighting}
\end{equation}
and $P_e=P_e^\top\succ0$ is the weighting matrix for each agent's global-state estimation error. The matrix $\bar P$ assigns the same weight $P_e$ to each agent's estimation error $e_i$.

Applying the barrier certificate condition \eqref{eq:general-bc-condition} to the closed-loop dynamics \eqref{eq:closed_loop_dynamics}, define
\begin{equation}
\begin{aligned}
    g_b(x,e,w)\triangleq{}&
    \nabla b(x)^\top\bigl(A_{bk}x-Ee+w\bigr)\\
    &+\alpha_b\bigl(b(x)\bigr).
\end{aligned}
\label{eq:gb}
\end{equation}
Assume that the extended class-$\mathcal K$ function $\alpha_b$ is chosen such that $\alpha_b(b(x))$ is polynomial in $x$.

\subsection{Sufficient Conditions for State Safety}
\label{subsec:sos-robust-bc}

We use sum-of-squares (SOS) conditions to establish a sufficient
condition for state safety. Let $\Phi(e)\triangleq1-V_e(e)$.
The first SOS condition certifies that $g_b(x,e,w)\geq0$ for all
$x\in\mathcal B$, $w\in\mathcal W$, and $e$ satisfying
$\Phi(e)\geq0$. The second certifies $\mathcal B\subseteq\mathcal C$.
State safety follows if the estimation error satisfies
$\Phi(e(t))\geq0$ for all $t\geq0$.

A real polynomial $p(z)$ in $z\in\mathbb R^d$ is an SOS polynomial
if there exist $r\in\mathbb N$ and real polynomials
$q_1,\ldots,q_r$ such that
$p(z)=\sum_{k=1}^{r}q_k(z)^2$. Thus, $p(z)\geq0$ for all $z\in\mathbb R^d$. Let $\Sigma[\cdot]$ denote the cone of SOS polynomials in the indicated variables.

\begin{lemma}
\label{lem:robust-bc}
For any initial state $x(0)\in\mathcal B$, let $g_b$
be defined by \eqref{eq:gb}. If $\Phi(e(t))\geq0$
for all $t\geq0$ and
\begin{equation}
    g_b-s_b b-s_\Phi\Phi-s_w\bigl(1-w^\top Q_w w\bigr)
    \in\Sigma[x,e,w]
    \label{eq:sos-robust-bc}
\end{equation}
for some SOS multipliers $s_b,s_\Phi,s_w\in\Sigma[x,e,w]$,
then $\mathcal B$ is forward invariant under
\eqref{eq:closed_loop_dynamics} for every disturbance satisfying
$w(t)\in\mathcal W$ for all $t\geq0$.
\end{lemma}

\begin{proof}
For $x\in\mathcal B$, \eqref{eq:barrier-set} gives
$b(x)\geq0$. For $w\in\mathcal W$,
\eqref{eq:disturbance-set} gives $1-w^\top Q_w w\geq0$.
Since $s_b$, $s_\Phi$, and $s_w$ are SOS,
\eqref{eq:sos-robust-bc} gives the following inequality
for all $x\in\mathcal B$, $w\in\mathcal W$, and
$e\in\mathbb R^{Nn}$ satisfying $\Phi(e)\geq0$:
\begin{equation}
\label{eq:bc-closedloop}
\begin{aligned}
g_b(x,e,w)
&\geq s_b b(x)+s_\Phi\Phi(e)\\
&\quad+s_w(1-w^\top Q_w w)\geq0.
\end{aligned}
\end{equation}
By \eqref{eq:general-bc-condition}, for any
$x(0)\in\mathcal B$, $x(t)\in\mathcal B$ for all $t\geq0$.
Hence, $\mathcal B$ is forward invariant. This completes our proof.
\end{proof}

Lemma~\ref{lem:robust-bc} requires $\Phi(e(t))\geq0$
for all $t\geq0$ to guarantee forward invariance of
$\mathcal B$. We therefore define the error-safe set
\begin{equation}
    \Omega\triangleq
    \{e\in\mathbb R^{Nn}:\Phi(e)\geq0\}.
    \label{eq:error-safe-set}
\end{equation}
Equivalently, $\Omega=\{e\in\mathbb R^{Nn}:V_e(e)\leq1\}$.

We next give an SOS condition for $\mathcal B\subseteq\mathcal C$ and combine it with the barrier condition to establish state safety.

\begin{proposition}
\label{prop:state-safety}
Let $0\leq t_1<t_2$. If $x(t_1)\in\mathcal B$,
$e(t)\in\Omega$ and $w(t)\in\mathcal W$ for all
$t\in[t_1,t_2]$, \eqref{eq:sos-robust-bc} holds, and there exists
$s_h\in\Sigma[x]$ such that
\begin{equation}
    h(x)-s_h(x)b(x)\in\Sigma[x],
    \label{eq:sos-safe-set-containment}
\end{equation}
then $x(t)\in\mathcal B\subseteq\mathcal C$ for all
$t\in[t_1,t_2]$.
\end{proposition}

\begin{proof}
Since $e(t)\in\Omega$ and $w(t)\in\mathcal W$ for all
$t\in[t_1,t_2]$, \eqref{eq:bc-closedloop} holds on
$[t_1,t_2]$. As $x(t_1)\in\mathcal B$,
\eqref{eq:general-bc-condition} gives $x(t)\in\mathcal B$
for all $t\in[t_1,t_2]$.

Moreover, \eqref{eq:sos-safe-set-containment} implies
$h(x)\geq s_h(x)b(x)$. Since $s_h$ is SOS, $b(x)\geq0$
implies $h(x)\geq0$, and hence $\mathcal B\subseteq\mathcal C$.
Therefore, $x(t)\in\mathcal B\subseteq\mathcal C$ for all
$t\in[t_1,t_2]$. This completes our proof.
\end{proof}

Lemma~\ref{lem:robust-bc} and Proposition~\ref{prop:state-safety} show that, if the SOS conditions~\eqref{eq:sos-robust-bc} and~\eqref{eq:sos-safe-set-containment} hold, maintaining $e(t)\in\Omega$ for all $t\geq0$ guarantees state safety. We formulate the problem as follows.
\begin{problem}
\label{prob:safety}
For the system in \eqref{eq:closed_loop_dynamics} with $x(0)\in\mathcal C$,
we find $P_e\in\mathbb S_{++}^n$, a polynomial barrier function
$b$ with $x(0)\in\mathcal B$, and SOS multipliers satisfying
\eqref{eq:sos-robust-bc} and \eqref{eq:sos-safe-set-containment}.
We then develop a decentralized topology-switching protocol
such that
\begin{enumerate}
\item[(i)] Each agent decides when to switch modes using only
its locally available information. For every $e(0)\in\Omega$,
the protocol maintains $V_e(e(t))\leq1$ and guarantees
$x(t)\in\mathcal B\subseteq\mathcal C$ for all $t\geq0$.
\item[(ii)] The protocol is non-Zeno. Each agent changes mode
only finitely many times on every bounded time interval.
\end{enumerate}
\end{problem}
\section{Topology-switching Protocol}
\label{sec:trigger}

In this section, we develop the decentralized topology-switching protocol in Problem~\ref{prob:safety}. By Lemma~\ref{lem:robust-bc} and Proposition~\ref{prop:state-safety}, maintaining $V_e(e(t))\leq1$ for all $t\geq0$ guarantees $x(t)\in\mathcal B\subseteq\mathcal C$ for all $t\geq0$ under the SOS conditions~\eqref{eq:sos-robust-bc} and~\eqref{eq:sos-safe-set-containment}. Since the state $x(t)$ is unknown, agent $i$ cannot compute $V_e(e(t))$ directly. We therefore construct a computable upper bound $U_i(t)$ that satisfies
$V_e(e(t))\leq U_i(t).$
Each agent in disconnected mode enters connected mode when $U_i(t)$ reaches $1$. Thus, at any time $t$ when there exists an agent $i$ in disconnected mode,
\begin{equation}
V_e(e(t))\leq U_i(t)\leq1,
\label{eq:disconnected-error-bound}
\end{equation}
so that $e(t)\in\Omega$. However, at any time $t$ when all agents are in connected mode, \eqref{eq:disconnected-error-bound} no longer applies. We therefore establish a recovery condition that makes $V_e$ decrease, i.e., $\dot V_e<0$ when $V_e\geq\gamma_{\mathrm{off}}$, so that $e(t)\in\Omega$. Here, $\gamma_{\mathrm{off}}\in(0,1)$ is a prescribed release threshold that determines when agents are ready to disconnect from the network.

Let $\Psi(t)\subseteq\mathcal N$ denote the set of agents in connected mode at time $t$. We construct $U_i(t)$ for each agent $i\in\mathcal N$ in three cases:

\begin{subequations}\label{eq:Ui-definition}
1) \emph{Full connectivity}: If every agent is in connected mode, i.e., $\Psi(t)=\mathcal N$, then all agents share the bound
\begin{equation}
U_i(t)\triangleq Z_{\mathrm{fc}}(t).
\label{eq:Ui-fc}
\end{equation}

2) \emph{Partial connectivity for agent $i$}: If $\Psi(t)\neq\mathcal N$ and agent $i$ has active neighbors, i.e., $\mathcal N_i(t)\neq\varnothing$, then agent $i$ uses
\begin{equation}
U_i(t)\triangleq Z_{\mathrm{pc},i}(t).
\label{eq:Ui-pc}
\end{equation}

3) \emph{No active neighbors for agent $i$}: If $\Psi(t)\neq\mathcal N$ and agent $i$ has no active neighbors, i.e., $\mathcal N_i(t)=\varnothing$, then agent $i$ uses
\begin{equation}
U_i(t)\triangleq Z_{\mathrm{dc},i}(t).
\label{eq:Ui-dc}
\end{equation}
This case applies to two scenarios: (i) an agent in disconnected mode and (ii) an agent in connected mode without active neighbors.
\end{subequations}

In Section~\ref{subsec:full-connectivity-certificate}, we establish a recovery condition under full connectivity and derive $Z_{\mathrm{fc}}$. In Section~\ref{subsec:partial-connectivity-bounds}, we derive $Z_{\mathrm{dc},i}$ and $Z_{\mathrm{pc},i}$. We then use the bounds derived in Sections~\ref{subsec:full-connectivity-certificate} and~\ref{subsec:partial-connectivity-bounds} to develop the topology-switching protocol in Section~\ref{subsec:topology-switching-protocol} and prove non-Zeno behavior and safety under the protocol in Section~\ref{subsec:non-zeno}.

\subsection{Recovery Condition and Error Bound under Full Connectivity}
\label{subsec:full-connectivity-certificate}

Under full connectivity, i.e., Case 1 in~\eqref{eq:Ui-fc}, we establish a recovery condition ensuring $\dot V_e<0$ when $V_e\geq\gamma_{\mathrm{off}}$ and derive $Z_{\mathrm{fc}}$. We then determine the recovery time $\tau_{\mathrm{rec}}$, i.e., the time required for $Z_{\mathrm{fc}}$ to decrease from $1$ to $\gamma_{\mathrm{off}}$. We also derive a uniform positive lower bound $\underline{\tau}_{\mathrm{rec}}$ satisfying $\tau_{\mathrm{rec}}\geq\underline{\tau}_{\mathrm{rec}}>0$ for the non-Zeno analysis in Section~\ref{subsec:non-zeno}.

Let $t_{\mathrm{fc}}$ denote the time when full connectivity begins. Let $t_{\mathrm{off}}$ denote the time when all agents simultaneously enter disconnected mode. We have $t_{\mathrm{off}}>t_{\mathrm{fc}}$. If $\Psi(t)=\mathcal N$ for all $t\in[t_{\mathrm{fc}},t_{\mathrm{off}})$, we call $[t_{\mathrm{fc}},t_{\mathrm{off}})$ a \emph{full-connectivity interval}. On this interval, $\mathcal G(t)$ is a connected spanning subgraph of $\bar{\mathcal G}$ with fixed Laplacian $\mathcal L(t)=\mathcal L_{\mathrm{fc}}$ for all $t\in[t_{\mathrm{fc}},t_{\mathrm{off}})$. We call \emph{$\mathcal L_{\mathrm{fc}}$ admissible} if $0\preceq\mathcal L_{\mathrm{fc}}\preceq\bar{\mathcal L}$ and $\lambda_2(\mathcal L_{\mathrm{fc}})>0$. The condition $\lambda_2(\mathcal L_{\mathrm{fc}})>0$ implies that all agents in full connectivity are connected through paths of active communication edges.

\subsubsection{Recovery Condition}
\label{subsubsec:recovery-condition}
On a full-connectivity interval $[t_{\mathrm{fc}},t_{\mathrm{off}})$, the error dynamics~\eqref{eq:dynamics-error} become
\begin{equation}
\dot e(t)=A_e(\mathcal L_{\mathrm{fc}})e(t)+\mathcal Iw(t)-J(\mathcal L_{\mathrm{fc}})v(t),
\label{eq:error-fc}
\end{equation}
for all $t\in[t_{\mathrm{fc}},t_{\mathrm{off}})$, where $A_e(\mathcal L_{\mathrm{fc}})=F(\mathcal L_{\mathrm{fc}})-\eta(\mathcal L_{\mathrm{fc}}\otimes I_n)$. Define $S_e(\mathcal L_{\mathrm{fc}})\triangleq A_e(\mathcal L_{\mathrm{fc}})^\top\bar P+\bar PA_e(\mathcal L_{\mathrm{fc}})$. We state the following proposition.

\begin{proposition}
\label{prop:recovery}
Let $V_e$ be defined as in~\eqref{eq:error-certificate} and satisfy $V_e(e(t))\geq\gamma_{\mathrm{off}}$ for all $t \in [t_{\mathrm{fc}},t_{\mathrm{off}})$.  For each admissible $\mathcal L_{\mathrm{fc}}$, given any $\bar P=I_N\otimes P_e\succ0$ and $\gamma_{\mathrm{off}}\in(0,1)$, if there exist scalars $\alpha\in\mathbb R_{>0}$ and $\beta_w,\beta_v\in\mathbb R_{\geq0}$ such that
\begin{subequations}
\label{eq:recovery-lmi}
\begin{equation}
\begin{bmatrix}
S_e(\mathcal L_{\mathrm{fc}})+\alpha\bar P
&\bar P\mathcal I&-\bar PJ(\mathcal L_{\mathrm{fc}})\\
\mathcal I^\top\bar P&-\beta_wQ_w&0\\
-J(\mathcal L_{\mathrm{fc}})^\top\bar P&0&-\beta_vR_v
\end{bmatrix}\preceq0,
\label{eq:full-connectivity-recovery}
\end{equation}
\begin{equation}
\beta_w+\beta_v<\alpha\gamma_{\mathrm{off}},
\label{eq:recovery-lmi-scalar}
\end{equation}
\end{subequations}
then $\dot V_e(e(t))<0$ for all $t \in [t_{\mathrm{fc}},t_{\mathrm{off}})$. 
% the error dynamics~\eqref{eq:error-fc} imply,
\end{proposition}

\begin{proof}
By~\eqref{eq:full-connectivity-recovery},
\[
\resizebox{0.98\columnwidth}{!}{$\displaystyle
\begin{bmatrix}
e(t)\\w(t)\\v(t)
\end{bmatrix}^{\!\top}
\begin{bmatrix}
S_e(\mathcal L_{\mathrm{fc}})+\alpha\bar P
&\bar P\mathcal I&-\bar PJ(\mathcal L_{\mathrm{fc}})\\
\mathcal I^\top\bar P&-\beta_wQ_w&0\\
-J(\mathcal L_{\mathrm{fc}})^\top\bar P&0&-\beta_vR_v
\end{bmatrix}
\begin{bmatrix}
e(t)\\w(t)\\v(t)
\end{bmatrix}
\leq0.
$}
\]
Expanding this inequality and using~\eqref{eq:error-fc} gives
$\dot V_e(e(t))\leq-\alpha V_e(e(t))
+\beta_w w(t)^\top Q_w w(t)
+\beta_v v(t)^\top R_vv(t)$ for all $t\in[t_{\mathrm{fc}},t_{\mathrm{off}})$.

Since $w(t)^\top Q_w w(t)\leq1$, $v(t)^\top R_vv(t)\leq1$ and $\beta_w,\beta_v\in\mathbb R_{\geq0}$ for all $t\in[t_{\mathrm{fc}},t_{\mathrm{off}})$, it follows that \begin{equation}
\dot V_e(e(t))\leq-\alpha V_e(e(t))+\beta_w+\beta_v.
\label{eq:full-connectivity-decay}
\end{equation} For $V_e(e(t))\geq\gamma_{\mathrm{off}}$,~\eqref{eq:recovery-lmi-scalar} gives $\dot V_e(e(t))\leq-\alpha\gamma_{\mathrm{off}}+\beta_w+\beta_v<0$. This completes our proof.
\end{proof}

\subsubsection{Error Bound Under Full-Connectivity and Recovery Time}
\label{subsubsec:common-bound-recovery-time}

For each admissible $\mathcal L_{\mathrm{fc}}$, let $\alpha$, $\beta_w$, and $\beta_v$ satisfy~\eqref{eq:full-connectivity-recovery} and~\eqref{eq:recovery-lmi-scalar}. These scalars depend on $\mathcal L_{\mathrm{fc}}$ and remain fixed within each full-connectivity interval. Set $\gamma_{\min}\triangleq(\beta_w+\beta_v)/\alpha\in[0,\gamma_{\mathrm{off}})$. Initialize $Z_{\mathrm{fc}}(t_{\mathrm{fc}})=1$ and let $\dot Z_{\mathrm{fc}}=-\alpha Z_{\mathrm{fc}}+\beta_w+\beta_v$, then $Z_{\mathrm{fc}}$ takes the following form:
\begin{equation}
Z_{\mathrm{fc}}(t)\triangleq \gamma_{\min}+(1-\gamma_{\min})e^{-\alpha(t-t_{\mathrm{fc}})}.
\label{eq:full-connectivity-certificate}
\end{equation}
We further define
\begin{equation}
\tau_{\mathrm{rec}}\triangleq
\frac1\alpha\log\!\left(
\frac{1-\gamma_{\min}}{\gamma_{\mathrm{off}}-\gamma_{\min}}
\right),
\label{eq:full-connectivity-recovery-time}
\end{equation}
and $t_{\mathrm{off}}\triangleq t_{\mathrm{fc}}+\tau_{\mathrm{rec}}$. Thus, $Z_{\mathrm{fc}}(t_{\mathrm{off}})=\gamma_{\mathrm{off}}$. We state the following lemma.

\begin{lemma}
\label{lem:full-connectivity-common-bound}
If $V_e(e(t_{\mathrm{fc}}))\leq1$, then $V_e(e(t))\leq Z_{\mathrm{fc}}(t)\leq1$ for all $t\in[t_{\mathrm{fc}},t_{\mathrm{off}}]$, and $V_e(e(t_{\mathrm{off}}))\leq\gamma_{\mathrm{off}}$.
\end{lemma}

\begin{proof}
Let $d(t)\triangleq V_e(e(t))-Z_{\mathrm{fc}}(t)$. By~\eqref{eq:full-connectivity-decay}, $\dot d\leq-\alpha d$, so $\frac{d}{dt}[e^{\alpha(t-t_{\mathrm{fc}})}d(t)]\leq0$. Since $d(t_{\mathrm{fc}})\leq0$, we obtain $d(t)\leq e^{-\alpha(t-t_{\mathrm{fc}})}d(t_{\mathrm{fc}})\leq0$. Equations~\eqref{eq:full-connectivity-certificate} and~\eqref{eq:full-connectivity-recovery-time} give $\gamma_{\mathrm{off}}\leq Z_{\mathrm{fc}}(t)\leq1$ and $Z_{\mathrm{fc}}(t_{\mathrm{off}})=\gamma_{\mathrm{off}}$. By continuity of $e$, $V_e(e(t_{\mathrm{off}}))\leq Z_{\mathrm{fc}}(t_{\mathrm{off}})=\gamma_{\mathrm{off}}$. This completes our proof.
\end{proof}

We next derive a positive lower bound $\underline\tau_{\mathrm{rec}}$ on $\tau_{\mathrm{rec}}$ that holds for all admissible $\mathcal L_{\mathrm{fc}}$. Given the observer gain $\bar L_k(\ell_{kk})$ in~\eqref{mode-dependent gain} for each $k\in\mathcal N$, we define
\begin{multline*}
a_k\triangleq-\operatorname{tr}(A_{bk}-B_kK_k)\\
+\max_{\bar L_k(\ell_{kk})\in\{\hat L_k,NL_k\}}
\operatorname{tr}\!\left(\bar L_k(\ell_{kk})C_k\right),
\end{multline*}
and set
\begin{equation}
\bar\alpha\triangleq
\frac{2}{Nn}
\left(\sum_{k\in\mathcal N}a_k+\eta n\,\operatorname{tr}(\bar{\mathcal L})\right).
\label{alpha_bar}
\end{equation}

\begin{lemma}
\label{lem:recovery-time-lower-bound}
For any admissible $\mathcal L_{\mathrm{fc}}$, if~\eqref{eq:full-connectivity-recovery} and~\eqref{eq:recovery-lmi-scalar} hold, then $0<\alpha\leq\bar\alpha$ and
\begin{equation}
\tau_{\mathrm{rec}}\geq
\underline\tau_{\mathrm{rec}}
\triangleq\frac1{\bar\alpha}\log\frac1{\gamma_{\mathrm{off}}}>0.
\label{eq:recovery-time-lower-bound}
\end{equation}
\end{lemma}

\begin{proof}
By~\eqref{eq:full-connectivity-recovery}, $S_e(\mathcal L_{\mathrm{fc}})+\alpha\bar P\preceq0$. Premultiplying and postmultiplying by $\bar P^{-1/2}$ gives
$\bar P^{-1/2}S_e(\mathcal L_{\mathrm{fc}})\bar P^{-1/2}
+\alpha I_{Nn}\preceq0.$ Taking the trace, substituting the definition of $S_e$, and using $\operatorname{tr}(XY)=\operatorname{tr}(YX)$ yields
\[
\begin{aligned}
0
&\geq\operatorname{tr}\!\left(
\bar P^{-1/2}S_e(\mathcal L_{\mathrm{fc}})\bar P^{-1/2}
\right)+Nn\,\alpha\\
&=\operatorname{tr}\!\left(
\bar P^{-1/2}A_e(\mathcal L_{\mathrm{fc}})^\top\bar P^{1/2}
\right)\\
&\quad+\operatorname{tr}\!\left(
\bar P^{1/2}A_e(\mathcal L_{\mathrm{fc}})\bar P^{-1/2}
\right)+Nn\,\alpha\\
&=2\operatorname{tr}(A_e(\mathcal L_{\mathrm{fc}}))+Nn\,\alpha.
\end{aligned}
\]
Here, $\operatorname{tr}(I_{Nn})=Nn$ and $\operatorname{tr}(A_e^\top)=\operatorname{tr}(A_e)$.
Since $\bar L_k(\ell_{kk})\in\{\hat L_k,NL_k\}$ for $k\in\mathcal N$, $\mathcal L_{\mathrm{fc}}\preceq\bar{\mathcal L}$, and $\eta\in\mathbb R_{\geq0}$,
\[
\begin{aligned}
&-\operatorname{tr}(A_e(\mathcal L_{\mathrm{fc}}))\\
&\quad=-\sum_{k\in\mathcal N}
\operatorname{tr}(A_{bk}-B_kK_k)\\
&\qquad+\sum_{k\in\mathcal N}
\operatorname{tr}\!\left(\bar L_k(\ell_{kk})C_k\right)+\eta n\,\operatorname{tr}(\mathcal L_{\mathrm{fc}})\\
&\quad\leq\sum_{k\in\mathcal N}a_k
+\eta n\,\operatorname{tr}(\bar{\mathcal L})=\frac{Nn}{2}\bar\alpha.
\end{aligned}
\]
Thus, $0<\alpha\leq\bar\alpha$. Finally, $0\leq\gamma_{\min}<\gamma_{\mathrm{off}}<1$ gives
$\frac{1-\gamma_{\min}}{\gamma_{\mathrm{off}}-\gamma_{\min}}
\geq\frac1{\gamma_{\mathrm{off}}},$ which, together with~\eqref{eq:full-connectivity-recovery-time}, proves~\eqref{eq:recovery-time-lower-bound}. This completes our proof.
\end{proof}

If $V_e(e(t_{\mathrm{fc}}))\leq1$, then $V_e(e(t))\leq Z_{\mathrm{fc}}(t)$ for all $t\in[t_{\mathrm{fc}},t_{\mathrm{off}}]$. Over the full-connectivity interval, $Z_{\mathrm{fc}}$ decreases from $1$ to $\gamma_{\mathrm{off}}$ in time $\tau_{\mathrm{rec}}$. However, when some agents are in disconnected mode, i.e., $\Psi(t)\neq\mathcal N$, each agent does not have access to the full Laplacian $\mathcal L(t)$. In the next subsection, we consider $\Psi(t)\neq\mathcal N$ and derive upper bounds on $V_e(e(t))$ that each agent can compute from its locally available information.

\subsection{Error Bounds for Connected Mode under Partial Connectivity and Disconnected Mode}
\label{subsec:partial-connectivity-bounds}

We now consider Case 2 in~\eqref{eq:Ui-pc} and Case 3 in~\eqref{eq:Ui-dc}. We derive the bounds $Z_{\mathrm{pc},i}$ and $Z_{\mathrm{dc},i}$ for each agent $i\in\mathcal N$. We first bound how quickly the estimation error can grow and solve a differential equation based on this bound to obtain upper bounds on $V_e(e(t))$. This equation has two topology-dependent coefficients: one relates to the estimation error, and the other to process disturbance $w$ and measurement noise $v$. We then bound these coefficients using only agent $i$'s locally available information, without requiring knowledge of the full Laplacian $\mathcal L(t)$. Substituting the coefficient bounds into the differential equation gives $Z_{\mathrm{pc},i}$ and $Z_{\mathrm{dc},i}$.

\subsubsection{Error Growth and Construction of Upper Bounds}
\label{subsec:comparison-bound}
Define the error radius $r_e(t)\triangleq\sqrt{V_e(e(t))}=\|\bar P^{1/2}e(t)\|_2$. To bound the growth of $r_e(t)$, define the topology-dependent coefficients
\begin{equation}
\begin{aligned}
\mu(\mathcal L(t))\triangleq\lambda_{\max}\!\Bigl(
\bar P^{-1/2}\bigl[
&A_e(\mathcal L(t))^\top\bar P\\
&+\bar P A_e(\mathcal L(t))
\bigr]\bar P^{-1/2}
\Bigr),
\end{aligned}
\label{eq:mu-definition}
\end{equation}
\begin{equation}
c(\mathcal L(t))\triangleq
\|\bar P^{1/2}\mathcal I Q_w^{-1/2}\|_2
+\|\bar P^{1/2}J(\mathcal L(t))R_v^{-1/2}\|_2.
\label{eq:c-definition}
\end{equation}
The following lemma bounds the growth of $r_e(t)$ while $\mathcal L(t)$ remains fixed.

\begin{lemma}
\label{lem:error-comparison-bound}
Let $t_3<t_4$ and $\mathcal L(t)=\mathcal L(t_3)=\mathcal L_0$ for all $t\in[t_3,t_4)$, we have that
\begin{equation}
D^+r_e(t)\leq\frac12\mu(\mathcal L_0)r_e(t)+c(\mathcal L_0),
~\forall t\in[t_3,t_4),
\label{eq:error-radius-differential-bound}
\end{equation}
where $D^+$ denotes the upper-right Dini derivative. 

The proof is given in Appendix~\ref{app:error-comparison-bound}.
\end{lemma}

However, when $\Psi(t)\neq\mathcal N$, agent $i$ cannot evaluate $\mu(\mathcal L(t))$ or $c(\mathcal L(t))$ because it has no access to the full Laplacian $\mathcal L(t)$. Suppose agent $i$ can compute upper bounds $\bar\mu_i\geq\mu(\mathcal L(t))$ and $\bar c_i\geq c(\mathcal L(t))$ from the topology of its connected component. The following lemma constructs $U_i(t)$ from these bounds. We then derive expressions for $\bar\mu_i$ and $\bar c_i$ in Section~\ref{subsec:topology-dependent-coefficients}.

\begin{lemma}
\label{lem:comparison-radius}
For each $i\in\mathcal N$, let $0\leq t_5<t_6$, $\Psi(t)\neq\mathcal N$ and $\mathcal L(t)=\mathcal L(t_5)=\mathcal L_1$ for all $t\in[t_5,t_6)$. Let $\bar\mu_i\in\mathbb R$ and $\bar c_i\in\mathbb R_{\geq0}$ satisfy
\begin{equation}
\mu(\mathcal L_1)\leq\bar\mu_i,
\quad
c(\mathcal L_1)\leq\bar c_i.
\label{eq:comparison-coefficient-bounds}
\end{equation}
Define $\bar r_{e,i}(t,\bar\mu_i,\bar c_i)$ for $t\in[t_5,t_6)$ by
\begin{equation}
\resizebox{0.84\columnwidth}{!}{$\displaystyle
\begin{cases}
e^{\frac12\bar\mu_i(t-t_5)}
\left(\bar r_{e,i}(t_5,\bar\mu_i,\bar c_i)+\dfrac{2\bar c_i}{\bar\mu_i}\right)
-\dfrac{2\bar c_i}{\bar\mu_i},
&\bar\mu_i\neq0,\\[2mm]
\bar r_{e,i}(t_5,\bar\mu_i,\bar c_i)+\bar c_i(t-t_5),
&\bar\mu_i=0.
\end{cases}$}
\label{eq:comparison-radius-solution}
\end{equation}
If $\bar r_{e,i}(t_5,\bar\mu_i,\bar c_i)\geq r_e(t_5)$, then, for all $t\in[t_5,t_6)$,
\begin{equation}
r_e(t)\leq\bar r_{e,i}(t,\bar\mu_i,\bar c_i),
\label{eq:comparison-radius-bound}
\end{equation}
and $V_e(e(t))\leq U_i(t)\triangleq\bar r_{e,i}^2(t,\bar\mu_i,\bar c_i)$.
\end{lemma}

\begin{proof}
By Lemma~\ref{lem:error-comparison-bound},~\eqref{eq:comparison-coefficient-bounds}, and $r_e(t)\geq0$, we have $D^+r_e(t)\leq\frac12\mu(\mathcal L_1)r_e(t)+c(\mathcal L_1)\leq\frac12\bar\mu_i r_e(t)+\bar c_i$. Solving $\dot{\bar r}_{e,i}(t,\bar\mu_i,\bar c_i)=\frac12\bar\mu_i\bar r_{e,i}(t,\bar\mu_i,\bar c_i)+\bar c_i$ with the given initial value yields~\eqref{eq:comparison-radius-solution}. Since $\bar r_{e,i}(t_5,\bar\mu_i,\bar c_i)\geq r_e(t_5)$, the comparison principle gives $0\leq r_e(t)\leq\bar r_{e,i}(t,\bar\mu_i,\bar c_i)$ for $t\in[t_5,t_6)$. Thus, $V_e(e(t))=r_e^2(t)\leq\bar r_{e,i}^2(t,\bar\mu_i,\bar c_i)=U_i(t)$ for $t\in[t_5,t_6)$. This completes our proof.
\end{proof}

\subsubsection{Topology-Dependent Coefficients}
\label{subsec:topology-dependent-coefficients}

We now derive $\bar\mu_i$ and $\bar c_i$ using only agent $i$'s locally available information. For agent $i$ in connected mode, let $\mathcal R_i(t)$ denote the set of agents in its connected component, consisting of $i$ and all agents reachable from $i$ through paths of active communication edges. For agent $i$ in disconnected mode, set $\mathcal R_i(t)=\varnothing$. We assume that agent $i$ has access to $\mathcal R_i(t)$ and the active communication edges between agents in $\mathcal R_i(t)$.

To bound $\mu(\mathcal L(t))$, we substitute $A_e(\mathcal L(t))=F(\mathcal L(t))-\eta(\mathcal L(t)\otimes I_n)$ into~\eqref{eq:mu-definition}. By~\eqref{Fij}, for any $i,j\in\mathcal N$,
\[
F_{ij}(\mathcal L(t))\triangleq
\scalebox{0.9}{$
\begin{cases}
A_{bi}-\bar L_i(\ell_{ii}(t))C_i-B_iK_i,
& \text{if } i=j,\\[4pt]
-B_jK_j,
& \text{if } i\neq j.
\end{cases}$}
\]
Recall $\bar P$ from~\eqref{eq:global-error-weighting}. Since $\mathcal L(t)=\mathcal L(t)^\top$, we have
\begin{equation}
\begin{aligned}
&\bar P^{-1/2}\bigl[A_e(\mathcal L(t))^\top\bar P
+\bar PA_e(\mathcal L(t))\bigr]\bar P^{-1/2}\\
&\quad=
\underbrace{
\bar P^{-1/2}\bigl[F(\mathcal L(t))^\top\bar P
+\bar PF(\mathcal L(t))\bigr]\bar P^{-1/2}
}_{M(\mathcal L(t))}\\
&\qquad-2\eta(\mathcal L(t)\otimes I_n).
\end{aligned}
\label{eq:normalized-F-matrix}
\end{equation}
Thus, $\mu(\mathcal L(t))=\lambda_{\max}\bigl(M(\mathcal L(t))-2\eta(\mathcal L(t)\otimes I_n)\bigr)$. Recall the observer gain defined in~\eqref{mode-dependent gain}. For each $k\in\mathcal N$, the gain follows one of two \emph{observer gain rules}:
\begin{enumerate}
\renewcommand{\labelenumi}{\arabic{enumi})}
\item \label{gain:no-active-neighbors}
\emph{Agent $k$ has no active neighbors}: If $\mathcal N_k(t)=\varnothing$, then $\ell_{kk}(t)=0$ and $\bar L_k(\ell_{kk}(t))=\hat L_k$.

\item \label{gain:active-neighbors}
\emph{Agent $k$ has at least one active neighbor}: If $\mathcal N_k(t)\neq\varnothing$, then $\ell_{kk}(t)>0$ and $\bar L_k(\ell_{kk}(t))=NL_k$.
\end{enumerate}
We then determine how the observer gains change $M(\mathcal L(t))$. Changing agent $k$'s observer gain affects only the $k$th diagonal block of both $F(\mathcal L(t))$ and $M(\mathcal L(t))$. Let $\Pi_k\triangleq[\,0_{n\times(k-1)n}\;\;I_n\;\;0_{n\times(N-k)n}\,]$, so that $\Pi_ke=e_k$. For any $j,k\in\mathcal N$, write $M_{jk}(\mathcal L(t))\triangleq\Pi_jM(\mathcal L(t))\Pi_k^\top$. In $M(0)$, every agent $k\in\mathcal N$ has observer gain $\hat L_k$. Replacing $\hat L_k$ by $NL_k$ changes $M_{kk}(0)$ by
\[
\begin{aligned}
\Delta M_k\triangleq-P_e^{-1/2}\bigl[
&C_k^\top(NL_k-\hat L_k)^\top P_e\\
&+P_e(NL_k-\hat L_k)C_k
\bigr]P_e^{-1/2}.
\end{aligned}
\]

We now define $\bar\mu_i$ and $\bar c_i$ for agent $i$, according to whether it has active neighbors:

\textbf{1) No active neighbors:}
By observer-gain rule~\ref{gain:no-active-neighbors}, agent $i$ has observer gain $\hat L_i$, if it is in connected mode without active neighbors or in disconnected mode. For any $k\in\mathcal N\setminus\{i\}$, the observer gain can be $\hat L_k$ or $NL_k$. The bound on $c(\mathcal L(t))$ takes the maximum over the norms corresponding to these possible observer gains. We define $(\bar\mu_i, \bar c_i) \triangleq (\bar\mu_i^{\mathrm{dis}}, \bar c_i^{\mathrm{dis}})$, where
\begin{equation}
\bar\mu_i^{\mathrm{dis}}\triangleq
\lambda_{\max}(M(0))
+\max\!\left\{0,\max_{k\in\mathcal N\setminus\{i\}}
\lambda_{\max}(\Delta M_k)\right\},
\label{eq:mu-dis-bound}
\end{equation}
\begin{equation}
\begin{aligned}
\bar c_i^{\mathrm{dis}}\triangleq{}&
\|\bar P^{1/2}\mathcal I Q_w^{-1/2}\|_2\\
&+\max\!\left\{
\begin{gathered}
\|P_e^{1/2}\hat L_iR_{v,i}^{-1/2}\|_2,\\
\displaystyle\max_{k\in\mathcal N\setminus\{i\}}
\|P_e^{1/2}\hat L_kR_{v,k}^{-1/2}\|_2,\\
\displaystyle\max_{k\in\mathcal N\setminus\{i\}}
\|P_e^{1/2}NL_kR_{v,k}^{-1/2}\|_2
\end{gathered}
\right\}.
\end{aligned}
\label{eq:c-dis-bound}
\end{equation}

\textbf{2) At least one active neighbor with $\Psi(t)\neq\mathcal N$:}
By observer-gain rule~\ref{gain:active-neighbors}, every agent $k\in\mathcal R_i(t)$ has observer gain $NL_k$, while each $k\in\mathcal N\setminus\mathcal R_i(t)$ can have observer gain $\hat L_k$ or $NL_k$. Let $\mathcal L_i^{\mathrm{known}}(t)\in\mathbb R^{N\times N}$ be the Laplacian formed from the active communication edges between agents in $\mathcal R_i(t)$. For any $k\in\mathcal N\setminus\mathcal R_i(t)$, the $k$th row and the $k$th column of $\mathcal L_i^{\mathrm{known}}(t)$ are zero. Define $(\bar\mu_i, \bar c_i) \triangleq (\bar\mu_i^{\mathrm{con}}, \bar c_i^{\mathrm{con}})$, where
\begin{equation}
\begin{aligned}
\bar\mu_i^{\mathrm{con}}(t)\triangleq{}&
\lambda_{\max}\!\Bigl(
M(0)-2\eta(\mathcal L_i^{\mathrm{known}}(t)\otimes I_n)\\
&\qquad+\sum_{k\in\mathcal R_i(t)}
\Pi_k^\top\Delta M_k\Pi_k
\Bigr)\\
&+\max\!\left\{0,
\max_{k\in\mathcal N\setminus\mathcal R_i(t)}
\lambda_{\max}(\Delta M_k)\right\},
\end{aligned}
\label{eq:mu-con-bound}
\end{equation}
\begin{equation}
\begin{aligned}
\bar c_i^{\mathrm{con}}(t)\triangleq{}&
\|\bar P^{1/2}\mathcal I Q_w^{-1/2}\|_2\\
&+\max\!\left\{
\begin{gathered}
\displaystyle\max_{k\in\mathcal R_i(t)}
\|P_e^{1/2}NL_kR_{v,k}^{-1/2}\|_2,\\
\displaystyle\max_{k\in\mathcal N\setminus\mathcal R_i(t)}
\|P_e^{1/2}\hat L_kR_{v,k}^{-1/2}\|_2,\\
\displaystyle\max_{k\in\mathcal N\setminus\mathcal R_i(t)}
\|P_e^{1/2}NL_kR_{v,k}^{-1/2}\|_2
\end{gathered}
\right\}.
\end{aligned}
\label{eq:c-con-bound}
\end{equation}

\begin{lemma}
\label{lem:local-comparison-coefficients}
Let $0\leq t_7<t_8$ such that $\Psi(t)\neq\mathcal N$ and
$\mathcal L(t)=\mathcal L(t_7)=\mathcal L_2$ for all
$t\in[t_7,t_8)$. For any $i\in\mathcal N$ and $t\in[t_7,t_8)$,
\begin{enumerate}
\item[1)] If $\mathcal N_i(t)=\varnothing$, agent $i$ follows observer-gain rule~\ref{gain:no-active-neighbors}. The coefficients defined in~\eqref{eq:mu-dis-bound} and~\eqref{eq:c-dis-bound} satisfy $\mu(\mathcal L_2)\leq\bar\mu_i^{\mathrm{dis}}$ and $c(\mathcal L_2)\leq\bar c_i^{\mathrm{dis}}$. The bound in~\eqref{eq:Ui-dc} is
\begin{equation}
U_i(t)=Z_{\mathrm{dc},i}(t)
\triangleq\bar r_{e,i}^2(t,\bar\mu_i^{\mathrm{dis}},\bar c_i^{\mathrm{dis}}).
\label{eq:Z-dc-definition}
\end{equation}

\item[2)] If $\mathcal N_i(t)\neq\varnothing$, agent $i$ follows observer-gain rule~\ref{gain:active-neighbors}. The coefficients defined in~\eqref{eq:mu-con-bound} and~\eqref{eq:c-con-bound} satisfy $\mu(\mathcal L_2)\leq\bar\mu_i^{\mathrm{con}}(t)$ and $c(\mathcal L_2)\leq\bar c_i^{\mathrm{con}}(t)$. The bound in~\eqref{eq:Ui-pc} is
\begin{equation}
U_i(t)=Z_{\mathrm{pc},i}(t)
\triangleq\bar r_{e,i}^2\bigl(t,\bar\mu_i^{\mathrm{con}}(t_7),
\bar c_i^{\mathrm{con}}(t_7)\bigr).
\label{eq:Z-pc-definition}
\end{equation}
\end{enumerate}
If $\bar r_{e,i}(t_7,\bar\mu_i,\bar c_i)\geq r_e(t_7)$, then
$V_e(e(t))\leq U_i(t)~\text{for all }t\in[t_7,t_8).$

The proof is given in Appendix~\ref{app:local-comparison-coefficients}.
\end{lemma}

When the topology changes at time $t_7$ with $\Psi(t_7^-)\neq\mathcal N$, the coefficients are updated according to~\eqref{eq:mu-dis-bound}, \eqref{eq:c-dis-bound}, \eqref{eq:mu-con-bound}, and~\eqref{eq:c-con-bound}. Let $t_8>t_7$ denote the next topology-switching time. If $\Psi(t)\neq\mathcal N$ for all $t\in[t_7,t_8)$, the radius $\bar r_{e,i}(t)$ is computed from~\eqref{eq:comparison-radius-solution} using the updated coefficients and the initial value $\bar r_{e,i}(t_7)=\bar r_{e,i}(t_7^-)$, where $\bar r_{e,i}(t)$ remains continuous at $t_7$.

Agent $i$ does not have access to the active edges in
$\mathcal L(t)-\mathcal L_i^{\mathrm{known}}(t)$. Each agent in this
unknown topology $\mathcal L(t)-\mathcal L_i^{\mathrm{known}}(t)$ may have one of two observer gains. Directly
checking all possible gain combinations can require up to
$2^{N-1}=O(2^N)$ topology-dependent calculations. The bounds
in~\eqref{eq:mu-dis-bound}, \eqref{eq:c-dis-bound},
\eqref{eq:mu-con-bound}, and~\eqref{eq:c-con-bound} avoid this
enumeration by taking maxima over agents' possible gains.

For each $k\in\mathcal N$, compute $\lambda_{\max}(\Delta M_k)$ for $\bar\mu_i$, and $\max\{0,\lambda_{\max}(\Delta M_k)\}$ bounds the largest eigenvalue of either possible diagonal-block change, $0$ or $\Delta M_k$. For $\bar c_i$, compute $\|P_e^{1/2}\hat L_kR_{v,k}^{-1/2}\|_2$ and $\|P_e^{1/2}NL_kR_{v,k}^{-1/2}\|_2$ for the two observer gains. Thus, the topology-dependent calculation of the scalar maxima in Lemma~\ref{lem:local-comparison-coefficients} uses $3N$ values, one eigenvalue and two norms per agent, and has complexity $O(N)$.

\subsection{Decentralized Topology-Switching Protocol}
\label{subsec:topology-switching-protocol}

We use the certificates in~\eqref{eq:full-connectivity-certificate}, \eqref{eq:Z-dc-definition}, and~\eqref{eq:Z-pc-definition} to trigger mode changes. We introduce a readiness bit $\sigma_i(t)$ so that agents in the same connected component can agree on when to disconnect, and a minimum waiting time $\underline\tau_{\mathrm{rec}}$ to prevent Zeno behavior. State safety still holds without the waiting time.

An agent $i$ in disconnected mode enters connected mode when
$Z_{\mathrm{dc},i}(t)$ reaches $1$. Under partial connectivity, an agent in connected mode is ready to disconnect when its certificate satisfies $U_i(t)\leq\gamma_{\mathrm{off}}$. Define the readiness bit as
\begin{equation}
\sigma_i(t)\triangleq
\begin{cases}
1, & \mathcal R_i(t)\neq\varnothing,\
     \mathcal R_i(t)\neq\mathcal N,\
     U_i(t)\leq\gamma_{\mathrm{off}},\\
0, & \text{otherwise}.
\end{cases}
\label{eq:readiness-bit}
\end{equation}
Here, $\sigma_i(t)=1$ indicates that agent $i$ is ready to disconnect, and $\sigma_i(t)=0$ indicates that it is not ready to disconnect. Each agent piggybacks $\sigma_i(t)$ on
routing or topology-control messages. We assume that, at each time
$t$, every agent in $\mathcal R_i(t)$ receives
$\{\sigma_j(t)\}_{j\in\mathcal R_i(t)}$ at that same time $t$.

When $\sigma_j(t)=1$ for all $j\in\mathcal R_i(t)$, the agents in
$\mathcal R_i(t)$ start a shared waiting interval. Let
$t_{\mathrm w}\in\mathbb R_{\geq0}\cup\{\varnothing\}$ denote the
start time of a waiting interval. The value
$t_{\mathrm w}=\varnothing$ means that no waiting interval is in
progress. If the membership of $\mathcal R_i(t)$ changes or any
readiness bit becomes $0$, the agents cancel the waiting interval
and set $t_{\mathrm w}=\varnothing$. Otherwise, all agents in
$\mathcal R_i(t)$ enter disconnected mode simultaneously when
$t-t_{\mathrm w}=\underline{\tau}_{\mathrm{rec}}$, where
$\underline{\tau}_{\mathrm{rec}}$ is defined
in~\eqref{eq:recovery-time-lower-bound}. After cancellation, a new waiting interval starts when all agents in the current connected component $\mathcal R_i(t)$ are ready.

If full connectivity begins at $t_{\mathrm{fc}}$, all waiting
intervals are canceled and every agent uses $Z_{\mathrm{fc}}(t)$,
initialized at $Z_{\mathrm{fc}}(t_{\mathrm{fc}})=1$. All agents remain
in connected mode with $\mathcal L(t)=\mathcal L_{\mathrm{fc}}$ for
$t\in[t_{\mathrm{fc}},t_{\mathrm{fc}}+\tau_{\mathrm{rec}})$, where
$\tau_{\mathrm{rec}}$ is given
by~\eqref{eq:full-connectivity-recovery-time}. At
$t_{\mathrm{off}}\triangleq t_{\mathrm{fc}}+\tau_{\mathrm{rec}}$,
$Z_{\mathrm{fc}}(t_{\mathrm{off}})=\gamma_{\mathrm{off}}$, and all agents
then enter disconnected mode simultaneously with
$\bar r_{e,i}(t_{\mathrm{off}}^+)=\sqrt{\gamma_{\mathrm{off}}}$.

Algorithm~\ref{alg:topology-switching-protocol} summarizes the
switching rules for agent $i$. Initially, set
$\bar r_{e,i}(0)=1$ and $t_{\mathrm w}=\varnothing$. Since
$e(0)\in\Omega$ implies $r_e(0)\leq1$, this initialization gives
$\bar r_{e,i}(0)\geq r_e(0)$ without requiring agent $i$ to know the
estimation error. We assume that agent $i$ continuously computes $U_i(t)$ and knows $\mathcal R_i(t)$ and the active communication edges between agents in $\mathcal R_i(t)$.

\begin{algorithm}[!tbp]
\caption{Topology-switching protocol for agent $i$}
\label{alg:topology-switching-protocol}
\DontPrintSemicolon

Assume $e(0)\in\Omega$.\;
\KwIn{Initial mode
$\operatorname{mode}_i(0)\in\{\mathrm{DM},\mathrm{CM}\}$, where
$\operatorname{mode}_i(t)$ denotes agent $i$'s mode at time $t$.}
\KwOut{Mode-switching decisions and certificate updates for agent $i$.}

$\bar r_{e,i}(0)\gets1,\quad
t_{\mathrm w}\gets\varnothing,\quad
\sigma_i(0)\gets0,\quad
t_{\mathrm{fc}}\gets\varnothing$\;

\For{$t\geq0$}{
    \uIf{$t_{\mathrm{fc}}\neq\varnothing$ or
    $\mathcal R_i(t)=\mathcal N$}{
        \If{$\mathcal R_i(t)=\mathcal N$ and
        $\bigl(t=0$ or $\mathcal R_i(t^-)\neq\mathcal N\bigr)$}{
            $t_{\mathrm{fc}}\gets t,\quad
            Z_{\mathrm{fc}}(t_{\mathrm{fc}})\gets1$\;
            $t_{\mathrm w}\gets\varnothing,\quad
            \sigma_i(t)\gets0$\;
        }
        Compute $U_i(t)=Z_{\mathrm{fc}}(t)$
        by~\eqref{eq:full-connectivity-certificate}\;
        \If{$U_i(t)\leq\gamma_{\mathrm{off}}$}{
            $\operatorname{mode}_i(t^+)\gets\mathrm{DM}$\;
            $\bar r_{e,i}(t^+)\gets\sqrt{\gamma_{\mathrm{off}}},~
            \sigma_i(t^+)\gets0,~
            t_{\mathrm{fc}}\gets\varnothing$\;
        }
    }
    \Else{
        \eIf{$\mathcal N_i(t)=\varnothing$}{
            Compute $U_i(t)=Z_{\mathrm{dc},i}(t)$
            by~\eqref{eq:Z-dc-definition}\;
        }{
            Compute $U_i(t)=Z_{\mathrm{pc},i}(t)$
            by~\eqref{eq:Z-pc-definition}\;
        }

        \uIf{$\operatorname{mode}_i(t)=\mathrm{DM}$}{
            \If{$U_i(t)\geq1$}{
                $\operatorname{mode}_i(t^+)\gets\mathrm{CM}$\;
            }
        }
        \Else{
            Update $\sigma_i(t)$ by~\eqref{eq:readiness-bit}\;

            \If{$t_{\mathrm w}\neq\varnothing$ and
            $\bigl(\mathcal R_i(t)\neq\mathcal R_i(t_{\mathrm w})$
            or $\sigma_j(t)=0$ for some $j\in\mathcal R_i(t)\bigr)$}{
                $t_{\mathrm w}\gets\varnothing$\;
            }
            \uIf{$t_{\mathrm w}\neq\varnothing$ and
            $t-t_{\mathrm w}=\underline{\tau}_{\mathrm{rec}}$}{
                $\operatorname{mode}_i(t^+)\gets\mathrm{DM}$\;
                $t_{\mathrm w}\gets\varnothing,\quad
                \sigma_i(t^+)\gets0$\;
            }
            \ElseIf{$t_{\mathrm w}=\varnothing$ and
            $\sigma_j(t)=1$ for all $j\in\mathcal R_i(t)$}{
                $t_{\mathrm w}\gets t$\;
            }
        }
    }
}
\end{algorithm}

\subsection{Non-Zeno Behavior and Safety Analysis}
\label{subsec:non-zeno}

We first establish non-Zeno behavior using the positive lower bound~\eqref{eq:recovery-time-lower-bound}. We then combine the estimation-error bounds under the topology-switching protocol with the SOS conditions~\eqref{eq:sos-robust-bc} and~\eqref{eq:sos-safe-set-containment} to establish state safety.

\begin{lemma}[Non-Zeno behavior]
\label{lem:non-zeno}
Let $\gamma_{\mathrm{off}}\in(0,1)$ be given and consider the error dynamics~\eqref{eq:dynamics-error} under Algorithm~\ref{alg:topology-switching-protocol}. If~\eqref{eq:recovery-lmi} holds for the admissible $\mathcal L_{\mathrm{fc}}$ used on each full-connectivity interval, then each agent changes mode only finitely many times on every bounded time interval.
\end{lemma}

\begin{proof}
By Lemma~\ref{lem:recovery-time-lower-bound}, $0<\underline{\tau}_{\mathrm{rec}}\leq\tau_{\mathrm{rec}}$. Fix $i\in\mathcal N$. Let $t_{i,k}^{\mathrm{on}}$ denote the $k$th time agent $i$ enters connected mode and $t_{i,k}^{\mathrm{off}}$ its subsequent disconnection time, if one occurs.

Under partial connectivity, agent $i$ disconnects at $t_{i,k}^{\mathrm{off}}=t_{\mathrm w}+\underline{\tau}_{\mathrm{rec}}$. Since the waiting interval starts after agent $i$ enters connected mode, and any increase in $|\mathcal R_i(t)|$ or loss of readiness cancels the waiting interval, $t_{\mathrm w}\geq t_{i,k}^{\mathrm{on}}$. Under full connectivity, agent $i$ disconnects at $t_{i,k}^{\mathrm{off}}=t_{\mathrm{fc}}+\tau_{\mathrm{rec}}$, where $t_{\mathrm{fc}}\geq t_{i,k}^{\mathrm{on}}$. Thus, in both cases,
$t_{i,k}^{\mathrm{off}}-t_{i,k}^{\mathrm{on}}
\geq\underline{\tau}_{\mathrm{rec}}>0.$

Fix any $T>0$. Since the time periods $[t_{i,k}^{\mathrm{on}},t_{i,k}^{\mathrm{off}})$ are disjoint and each has duration at least $\underline{\tau}_{\mathrm{rec}}$, agent $i$ can enter disconnected mode at most $1+\lfloor T/\underline{\tau}_{\mathrm{rec}}\rfloor$ times on $[0,T]$, including the case where it is initially in connected mode. Since agent $i$ must enter connected mode before it can enter disconnected mode again, the number of times it enters connected mode is also finite. Since $\mathcal N$ is finite, the entire protocol is non-Zeno. This completes our proof.
\end{proof}

We next show that the protocol keeps $e(t)\in\Omega$ for all $t\geq0$ and thus guarantees state safety by Proposition~\ref{prop:state-safety}. Together with Lemma~\ref{lem:non-zeno}, this shows that the protocol satisfies both requirements of Problem~\ref{prob:safety}.

\begin{theorem}
\label{thm:safety-under-switching}
Let $x(0)\in\mathcal B$ and $e(0)\in\Omega$. If
\begin{itemize}
\item the SOS conditions~\eqref{eq:sos-robust-bc} and~\eqref{eq:sos-safe-set-containment} are satisfied, and
\item for each full-connectivity interval, there exist $\alpha\in\mathbb R_{>0}$ and $\beta_w,\beta_v\in\mathbb R_{\geq0}$ satisfying~\eqref{eq:full-connectivity-recovery} and~\eqref{eq:recovery-lmi-scalar} for the admissible $\mathcal L_{\mathrm{fc}}$ used on that interval,
\end{itemize}
then the topology-switching protocol in Algorithm~\ref{alg:topology-switching-protocol} has the following properties:
\begin{enumerate}
\item[(i)] The protocol maintains $V_e(e(t))\leq1$ for all $t\geq0$ and guarantees $x(t)\in\mathcal B\subseteq\mathcal C$ for all $t\geq0$, as required in Problem~\ref{prob:safety}(i).
\item[(ii)] The protocol is non-Zeno, as required in Problem~\ref{prob:safety}(ii).
\end{enumerate}
\end{theorem}

\begin{proof}
To prove (i), let $t_0=0$, and let $t_{k+1}$ be the next topology-change time after $t_k$, or $\infty$ if there is no further change. Write $\mathcal L(t)=\mathcal L(t_k)=\mathcal L_k$ for all $t\in[t_k,t_{k+1})$. We prove $V_e(e(t))\leq1$ on each interval. Initially, $e(0)\in\Omega$ gives $r_e(0)\leq1$, and each $\bar r_{e,i}$ is initialized at $1$.

On an interval with $\Psi(t)\neq\mathcal N$, the radius in~\eqref{eq:Z-dc-definition} or~\eqref{eq:Z-pc-definition} starts with $\bar r_{e,i}(t_k,\bar\mu_i,\bar c_i)\geq r_e(t_k)$. Lemma~\ref{lem:local-comparison-coefficients} gives $V_e(e(t))\leq U_i(t)$ for every $i\in\mathcal N$. At least one agent is in disconnected mode, and its connection rule gives $U_i(t)=Z_{\mathrm{dc},i}(t)\leq1$. Thus, $V_e(e(t))\leq1$. If $t_{k+1}<\infty$ and the next interval also has $\Psi(t)\neq\mathcal N$,~\eqref{eq:comparison-radius-bound} and continuity give $r_e(t_{k+1})\leq\bar r_{e,i}(t_{k+1}^-,\bar\mu_i,\bar c_i)$. The radius for the next interval is initialized at this left limit.

On an interval with $\Psi(t)=\mathcal N$ and $\mathcal L_k=\mathcal L_{\mathrm{fc}}$, $V_e$ is at most $1$ at its start. Lemma~\ref{lem:full-connectivity-common-bound} gives $V_e(e(t))\leq Z_{\mathrm{fc}}(t)\leq1$. At the release time $t_{\mathrm{off}}$, the protocol has $Z_{\mathrm{fc}}(t_{\mathrm{off}})\leq\gamma_{\mathrm{off}}$, so $r_e(t_{\mathrm{off}})\leq\sqrt{\gamma_{\mathrm{off}}}$. Algorithm~\ref{alg:topology-switching-protocol} initializes $\bar r_{e,i}(t_{\mathrm{off}}^+)=\sqrt{\gamma_{\mathrm{off}}}$. These initializations preserve $V_e(e(t))\leq U_i(t)$ across topology switches. Hence, $V_e(e(t))\leq1$ for all $t\geq0$.

The SOS conditions~\eqref{eq:sos-robust-bc} and~\eqref{eq:sos-safe-set-containment}, together with Proposition~\ref{prop:state-safety}, give $x(t)\in\mathcal B\subseteq\mathcal C$ for all $t\geq0$. This proves (i). Claim (ii) follows from Lemma~\ref{lem:non-zeno}. This completes our proof.
\end{proof}
\section{Computing Safety Certificates}
\label{sec:optimization}

With the controller and observer gains and $\mathcal L_{\mathrm{fc}}$ fixed, we formulate a problem to compute $P_e$, which defines $\bar P$ in~\eqref{eq:global-error-weighting}, a quadratic barrier function, the scalar variables in~\eqref{eq:recovery-lmi}, and the SOS multipliers in~\eqref{eq:sos-robust-bc} and~\eqref{eq:sos-safe-set-containment}. The problem minimizes a bound on estimation-error growth under full disconnection subject to the recovery and safety conditions. Since the constraints are bilinear, we adopt an alternating method~\cite{dai2023convex} to find a feasible solution satisfying the recovery and safety conditions. This solution establishes state safety under the topology-switching protocol.

\subsection{Bilinear Problem}
\label{subsec:joint-design-problem}

We choose $b(x)=1-x^\top H_bx$ with $H_b\in\mathbb S_{++}^n$, impose $\operatorname{tr}(P_e)=n$ to fix the scale of the error certificate, and set $\alpha_b(s)=\kappa_bs$ for a fixed $\kappa_b>0$. We restrict the multipliers $s_b,s_\Phi,s_w,s_h$ in~\eqref{eq:sos-robust-bc} and~\eqref{eq:sos-safe-set-containment} to non-negative scalars.

Under full disconnection with $w=v=0$ in~\eqref{system},~\eqref{eq:dynamics-error} gives $\dot V_e=e^\top(A_e(0)^\top\bar P+\bar PA_e(0))e$. Define
$J(P_e)\triangleq
\lambda_{\max}(A_e(0)^\top\bar P+\bar PA_e(0))$,
so that $\dot V_e\leq J(P_e)\|e\|_2^2$. We minimize this bound subject to the recovery and safety conditions.

Define $\mathcal X_s\triangleq\{\alpha,\beta_w,\beta_v,s_b,s_\Phi,s_w,s_h\}$ and $\mathcal X_c\triangleq\{P_e,H_b\}$. We formulate
\begin{equation}
\begin{aligned}
\underset{\mathcal X_c,\mathcal X_s}{\operatorname{minimize}}
\quad &J(P_e)\\
\operatorname{subject\ to}\quad
    &P_e,H_b\in\mathbb S_{++}^{n},\\
    &\operatorname{tr}(P_e)=n,\quad b(x(0))\geq0,\\
    &s_b,s_\Phi,s_w,s_h\geq0,\\
    &\eqref{eq:sos-robust-bc},\quad
      \eqref{eq:sos-safe-set-containment},\\
    &\alpha>0,\quad\beta_w,\beta_v\geq0,\\
    &\beta_w+\beta_v<\alpha,\\
    &\eqref{eq:full-connectivity-recovery}.
\end{aligned}
\label{eq:joint-design-problem}
\end{equation}
Here, $b(x(0))\geq0$ ensures $x(0)\in\mathcal B$, the SOS constraints certify the barrier condition and set containment, and~\eqref{eq:full-connectivity-recovery} imposes recovery. Any feasible solution has $\gamma_{\min}\triangleq(\beta_w+\beta_v)/\alpha<1$. Selecting $\gamma_{\mathrm{off}}\in(\gamma_{\min},1)$ then satisfies~\eqref{eq:recovery-lmi-scalar}.

\begin{lemma}
\label{lem:optimization-safety}
Let $(\mathcal X_c,\mathcal X_s)$ be feasible for~\eqref{eq:joint-design-problem}, and select $\gamma_{\mathrm{off}}\in(\gamma_{\min},1)$. Assume $e(0)\in\Omega$ and that the agents follow Algorithm~\ref{alg:topology-switching-protocol} with the fixed full-connectivity Laplacian $\mathcal L_{\mathrm{fc}}$. For every $w$ and $v$ satisfying $w(t)\in\mathcal W$ and $v(t)\in\mathcal V$ for all $t\geq0$, we have $e(t)\in\Omega$ and $x(t)\in\mathcal B\subseteq\mathcal C$ for all $t\geq0$.
\end{lemma}

\begin{proof}
Feasibility of~\eqref{eq:joint-design-problem} gives $b(x(0))\geq0$, the SOS conditions~\eqref{eq:sos-robust-bc} and~\eqref{eq:sos-safe-set-containment}, and~\eqref{eq:full-connectivity-recovery} for the fixed $\mathcal L_{\mathrm{fc}}$. Since $H_b\succ0$, $\nabla b(x)\neq0$ on $\partial\mathcal B$. Also, $\gamma_{\mathrm{off}}>\gamma_{\min}=(\beta_w+\beta_v)/\alpha$ gives~\eqref{eq:recovery-lmi-scalar}. Thus, Lemma~\ref{lem:non-zeno} and Theorem~\ref{thm:safety-under-switching} give the result.
\end{proof}

\subsection{Alternating Optimization}
\label{subsec:alternating-optimization}

Problem~\eqref{eq:joint-design-problem} contains the bilinear products $\alpha\bar P$, $s_\Phi\bar P$, $s_bH_b$, and $s_hH_b$. Each product contains one variable from $\mathcal X_c$ and one from $\mathcal X_s$. With the variables in either set fixed, optimization over the variables in the other set can be formulated as a semidefinite program (SDP). So, we optimize over these two sets, $\mathcal X_c~\text{and}~\mathcal X_s$, alternately. With $\mathcal X_s$ fixed, minimizing $J(P_e)$ is equivalent to minimizing a scalar $\lambda_{\mathrm{dis}}\in\mathbb R$ subject to 
\begin{equation}
A_e(0)^T\bar P+\bar P A_e(0)
\preceq \lambda_{\mathrm{dis}} I.
\label{eq:disconnection-objective-lmi}
\end{equation}
and the constraints in \eqref{eq:joint-design-problem}.

Starting from a feasible point $(\mathcal X_c^{(0)},\mathcal X_s^{(0)})$, we first fix $\mathcal X_s=\mathcal X_s^{(k)}$ and solve \eqref{eq:joint-design-problem} over $\mathcal X_c$ to obtain $\mathcal X_c^{(k+1)}$. We then fix $\mathcal X_c=\mathcal X_c^{(k+1)}$ and optimize over $\mathcal X_s$. Since $J(P_e)$ depends only on $P_e$, it remains constant while optimizing over $\mathcal X_s$. We therefore minimize the recovery ratio $\gamma_{\min}=(\beta_w+\beta_v)/\alpha$ subject to the same constraints. A release threshold $\gamma_{\mathrm{off}}$ can then be chosen in $(\gamma_{\min},1)$.

To minimize this ratio, we use bisection on an upper bound $\gamma$ and impose $\beta_w+\beta_v\leq\gamma\alpha$. For each fixed $\gamma$, this inequality and the constraints in \eqref{eq:joint-design-problem} define an SDP problem over $\mathcal X_s$. If a feasible solution exists for some $\gamma$, the same solution is feasible for every larger $\gamma$, which permits bisection.

After the matrix update, $(\mathcal X_c^{(k+1)},\mathcal X_s^{(k)})$ satisfies all constraints in \eqref{eq:joint-design-problem}. We therefore initialize $\underline\gamma=0$ and $\overline\gamma=(\beta_w^{(k)}+\beta_v^{(k)})/\alpha^{(k)}$, and set $\mathcal X_s^{(k+1)}=\mathcal X_s^{(k)}$. At each bisection step, set $\gamma=(\underline\gamma+\overline\gamma)/2$ and solve the SDP problem. If a feasible solution exists, set $\overline\gamma=\gamma$ and assign that solution to $\mathcal X_s^{(k+1)}$. Otherwise, set $\underline\gamma=\gamma$. Stop bisection when $\overline\gamma-\underline\gamma\leq\varepsilon_\gamma$, where $\varepsilon_\gamma>0$ is a prescribed tolerance.

Let $J^{(k)}\triangleq J(P_e^{(k)})$. The preceding matrices $\mathcal X_c^{(k)}$ remain feasible during the matrix update, so solving that update exactly gives $J^{(k+1)}\leq J^{(k)}$. The scalar update leaves $P_e^{(k+1)}$ unchanged and therefore does not change $J^{(k+1)}$. Moreover, $\operatorname{tr}(P_e)=n$ implies $0\prec P_e\preceq nI_n$, and hence $J(P_e)\geq-2n\|A_e(0)\|_2$. Thus, the objective values are nonincreasing and bounded below, so they converge.

Let $\theta\in\{(P_e)_{ij},(H_b)_{ij},\alpha,\beta_w,\beta_v,s_b,s_\Phi,s_w,s_h:\ i,j=1,\ldots,n\}$, and denote the value of $\theta$ at iteration $k$ by $\theta^{(k)}$. The stopping conditions are
\begin{equation}
\begin{aligned}
|J^{(k)}-J^{(k-1)}|&<\varepsilon_J,\\
|\theta^{(k)}-\theta^{(k-1)}|&<\varepsilon_X
\quad\text{for every }\theta,
\end{aligned}
\label{eq:alternating-stopping-conditions}
\end{equation}
or when $k=k_{\max}$. Here, $\varepsilon_J,\varepsilon_X>0$ are prescribed tolerances. Algorithm~\ref{alg:joint-certificate-computation} summarizes the procedure.

\begin{algorithm}[!tbp]
\caption{Alternating certificate computation}
\label{alg:joint-certificate-computation}
\DontPrintSemicolon

\KwIn{Feasible $(\mathcal X_c^{(0)},\mathcal X_s^{(0)})$;
$\varepsilon_\gamma,\varepsilon_J,\varepsilon_X>0$;
$k_{\max}\geq1$.}

$k\gets0,\quad J^{(0)}\gets J(P_e^{(0)})$\;

\Repeat{\eqref{eq:alternating-stopping-conditions} holds
or $k=k_{\max}$}{
    Fix $\mathcal X_s=\mathcal X_s^{(k)}$ and minimize
    $\lambda_{\mathrm{dis}}$ over $\mathcal X_c,\lambda_{\mathrm{dis}}$
    subject to~\eqref{eq:joint-design-problem}
    and~\eqref{eq:disconnection-objective-lmi};
    obtain $\mathcal X_c^{(k+1)}$\;

    Fix $\mathcal X_c=\mathcal X_c^{(k+1)}$\;
    $\underline\gamma\gets0,\quad
    \overline\gamma\gets
    (\beta_w^{(k)}+\beta_v^{(k)})/\alpha^{(k)}$\;
    $\mathcal X_s^{(k+1)}\gets\mathcal X_s^{(k)}$\;

    \While{$\overline\gamma-\underline\gamma>\varepsilon_\gamma$}{
        $\gamma\gets(\underline\gamma+\overline\gamma)/2$\;
        Find $\mathcal X_s$ satisfying
        \eqref{eq:joint-design-problem} and
        $\beta_w+\beta_v\leq\gamma\alpha$\;
        \eIf{a solution $\mathcal X_s^\star$ exists}{
            $\overline\gamma\gets\gamma,\quad
            \mathcal X_s^{(k+1)}\gets\mathcal X_s^\star$\;
        }{
            $\underline\gamma\gets\gamma$\;
        }
    }

    $J^{(k+1)}\gets J(P_e^{(k+1)}),\quad k\gets k+1$\;
}

$\gamma_{\min}\gets
(\beta_w^{(k)}+\beta_v^{(k)})/\alpha^{(k)}$\;
Select $\gamma_{\mathrm{off}}\in(\gamma_{\min},1)$\;
\KwRet{$\mathcal X_c^{(k)}$, $\mathcal X_s^{(k)}$,
$\gamma_{\mathrm{off}}$}\;
\end{algorithm}
\section{Case Study}
\label{sec:simulation}

We evaluate the topology-switching protocol in Algorithm~\ref{alg:topology-switching-protocol} on the four interconnected microgrids of case~(a2) in~\cite{9507261}. The power lines are fixed. Connected and disconnected modes refer only to the communication used to exchange state estimates. We evaluate safety under bounded disturbances and measurement noise and measure the reduction in communication.

\subsection{Microgrid Model and Equilibrium}
\label{subsec:test-system}
\label{subsec:equilibrium}

The power line admittances and microgrid parameters are listed in Tables~\ref{tab:microgrid-lines} and~\ref{tab:microgrid-parameters}. For each power line $(i,j)$, $g_{ij}+\mathrm{j}b_{ij}$ denotes its admittance, with $\mathrm{j}=\sqrt{-1}$, conductance $g_{ij}$, and susceptance $b_{ij}$. The coefficients $m_i=\tau_i/k_i$ and $d_i=1/k_i$ denote virtual inertia and damping, where $k_i$ is the droop gain and $\tau_i$ is the measurement-filter time constant. The quantities $P_i^s$, $V_i$, and $\theta_i^{\rm src}$ are the reported active-power setpoint, voltage magnitude, and phase. In case~(a2), $P_4^s$ changes from $-4.06$ to $-7.06$~p.u. and $P_3^s$ from $-2.25$ to $0.25$~p.u., which causes the resulting operating point unstable without feedback~\cite{9507261}.

Let $\theta_i$ and $\omega_i$ denote voltage phase and angular-frequency deviation. With feedback input $u_i$, constant correction $\bar u_i$, and acceleration disturbance $w_i^a$, the nonlinear dynamics are
\begin{equation}
\begin{aligned}
\dot\theta_i&=\omega_i,\\
m_i\dot\omega_i&=P_i^s+\bar u_i+u_i-P_i^e(\theta)-d_i\omega_i+m_iw_i^a,
\end{aligned}
\label{eq:sim-nonlinear-plant}
\end{equation}
where $\mathcal E_{\rm el}$ is the set of fixed power lines,
\begin{equation}
P_i^e(\theta)=\sum_{j:(i,j)\in\mathcal E_{\rm el}}\!\left[g_{ij}V_i^2-V_iV_j\bigl(g_{ij}\cos\theta_{ij}+b_{ij}\sin\theta_{ij}\bigr)\right],
\label{eq:sim-power-flow}
\end{equation}
with $\theta_{ij}=\theta_i-\theta_j$. Using $\theta_1^\star=0.47$~rad from~\cite{9507261}, we solve $P_i^e(\theta^\star)=P_i^s$ for $i=1,2,3$ and obtain $\theta^\star=[0.47000,\,0.06253,\,-0.25872,\,-0.38629]^T$~rad. Setting $\bar u=[0,\,0,\,0,\,0.048578]^T$~p.u. gives $P^e(\theta^\star)=P^s+\bar u$, so $(\theta^\star,0)$ is an equilibrium of~\eqref{eq:sim-nonlinear-plant} for $u=w^a=0$.

\makeatletter
\newcommand{\casestudysinglelinecaption}[1]{%
  \let\casestudyoldmakecaption\@makecaption
  \long\def\@makecaption##1##2{%
    \par{\centering\normalfont\footnotesize ##1.\ \scshape ##2\par}%
    \vspace{0.5\baselineskip}}%
  \caption{#1}%
  \let\@makecaption\casestudyoldmakecaption}
\makeatother

\begin{table}[!t]
\casestudysinglelinecaption{Electrical line admittances.}
\label{tab:microgrid-lines}
\vspace{-3pt}
\centering
\footnotesize
\setlength{\tabcolsep}{6pt}
\begin{tabular}{cc}
\hline
Line & $g_{ij}+\mathrm{j}b_{ij}$ (p.u.)\\
\hline
$1$--$2$ & $2.71-\mathrm{j}9.14$\\
$1$--$4$ & $8.47-\mathrm{j}9.77$\\
$2$--$3$ & $4.07-\mathrm{j}4.32$\\
$2$--$4$ & $0.64-\mathrm{j}5.17$\\
$3$--$4$ & $4.64-\mathrm{j}9.93$\\
\hline
\end{tabular}
\end{table}

\begin{table}[!t]
\casestudysinglelinecaption{Microgrid parameters for case~(a2).}
\label{tab:microgrid-parameters}
\vspace{-4pt}
\centering
\footnotesize
\setlength{\tabcolsep}{2.5pt}
\begin{tabular}{cccccc}
\hline
$i$ & $m_i$ (s) & $d_i$ & $P_i^s$ (p.u.) & $V_i$ (p.u.) & $\theta_i^{\rm src}$ (rad)\\
\hline
1 & 5.76 & 1.03 & 13.13 & 0.96 & 0.47\\
2 & 9.20 & 1.61 & 0.39 & 0.95 & 0.07\\
3 & 9.32 & 1.86 & 0.25 & 0.99 & $-0.25$\\
4 & 4.92 & 1.50 & $-7.06$ & 1.02 & $-0.37$\\
\hline
\end{tabular}
\end{table}

\subsection{Local Model and Disturbance Bound}
\label{subsec:linear-model}

We simulate the nonlinear plant~\eqref{eq:sim-nonlinear-plant}, while the controller and observers use its local linear model. Let $q=\theta-\theta^\star$, $r=\omega$, $M=\operatorname{diag}(m_1,\ldots,m_4)$, $D=\operatorname{diag}(d_1,\ldots,d_4)$, and $H_\theta=\left.\partial P^e/\partial\theta\right|_{\theta^\star}$. For the scaled state $x=10^2[q^T\;r^T]^T$, the dynamics have the form $\dot x=Ax+Bu+\widetilde w$, with $A=\left[\begin{smallmatrix}0&I_4\\-M^{-1}H_\theta&-M^{-1}D\end{smallmatrix}\right]$, $B=10^2[0\;M^{-1}]^T$, and $\widetilde w=10^2[0^T\;(w^a+d_{\rm nl}(q))^T]^T$, where $d_{\rm nl}(q)=-M^{-1}[P^e(\theta^\star+q)-P^e(\theta^\star)-H_\theta q]$.

Let $\rho>0$ (rad) bound the deviation $|q_i-q_j|$ of the phase difference across every power line $(i,j)\in\mathcal E_{\rm el}$. Under this condition, Taylor's theorem applied to~\eqref{eq:sim-power-flow} gives $|d_{{\rm nl},i}(q)|\le\bar d_{{\rm nl},i}(\rho)$, where
\begin{equation}
\begin{aligned}
\bar d_{{\rm nl},i}(\rho)=\frac{\rho^2}{2m_i}\sum_{j:(i,j)\in\mathcal E_{\rm el}}V_iV_j
\max_{|\xi-\theta_{ij}^\star|\le\rho}\bigl|g_{ij}\cos\xi+b_{ij}\sin\xi\bigr|.
\end{aligned}
\label{eq:sim-remainder-bound}
\end{equation}
Consequently, $\|d_{\rm nl}(q)\|_2\le\bar d_{\rm nl}(\rho)\triangleq(\sum_{i=1}^4\bar d_{{\rm nl},i}^2(\rho))^{1/2}$. With $\|w^a\|_2\le\bar w_a=10^{-6}$~rad/s$^2$, we choose $Q_w^{-1/2}(\rho)=\operatorname{diag}(10^{-6}I_4,\,10^2[\bar w_a+\bar d_{\rm nl}(\rho)]I_4)$ for the disturbance set $\mathcal W$ defined previously. The first four entries are strictly positive to keep $Q_w\succ0$, although the corresponding components of $\widetilde w$ are zero. The measurements are $y_i=10^2[q_i\;r_i]^T+v_i=C_ix+v_i$, with $R_v^{-1/2}=0.03I_8$, corresponding to componentwise noise bounds of $3\times10^{-4}$~rad and rad/s.

\subsection{Safety Design and Simulation Setup}
\label{subsec:safety-region}
\label{subsec:simulation-setup}

Recall~\eqref{eq:safe-region} and~\eqref{eq:barrier-set}. We require $|q_i|\le0.05$~rad and $|r_i|\le0.05$~rad/s for $i=1,\ldots,4$, defining $\mathcal C=\{x\in\mathbb R^8:|q_i|\le0.05,\ |r_i|\le0.05,\ i=1,\ldots,4\}$. For the results, define $S_q(t)=\max_i|q_i(t)|/(0.05\,\mathrm{rad})$, $S_r(t)=\max_i|r_i(t)|/(0.05\,\mathrm{rad/s})$, and $S(t)=\max\{S_q(t),S_r(t)\}$. Thus, $x(t)\in\mathcal C$ exactly when $S(t)\le1$.

The constraints defining $\mathcal C$ do not imply $|q_i-q_j|\le\rho$, which is required by~\eqref{eq:sim-remainder-bound}. We therefore set $\mathcal X_{\rm lin}(\rho)=\{x\in\mathcal C:|q_i-q_j|\le\rho,\ (i,j)\in\mathcal E_{\rm el}\}$ and choose a quadratic barrier certificate $b(x)=1-x^TH_bx$, $H_b\succ0$, with $\mathcal B=\{x:b(x)\ge0\}$. Requiring $\mathcal B\subseteq\mathcal X_{\rm lin}(\rho)$ ensures that the disturbance bound and the phase and frequency constraints hold throughout $\mathcal B$.

For the standard basis vectors $a_k\in\mathbb R^8$, the containment constraints take the common form $s_hH_b\succeq\zeta\zeta^T$, $0\le s_h\le1$, with $\zeta=a_i/5$ or $\zeta=a_{4+i}/5$ for $i=1,\ldots,4$, and $\zeta=(a_i-a_j)/(10^2\rho)$ for $(i,j)\in\mathcal E_{\rm el}$. We apply~\eqref{eq:sos-safe-set-containment} separately to each phase, frequency, and phase-difference constraint in $\mathcal X_{\mathrm{lin}}(\rho)$. Together with the barrier condition~\eqref{eq:sos-robust-bc} and the condition $V_e(e(t))\leq1$ maintained by Algorithm~\ref{alg:topology-switching-protocol}, Theorem~\ref{thm:safety-under-switching} applies to the nonlinear plant.

We use LQR with $Q_K=\operatorname{diag}(20I_4,I_4)$ and $R_K=I_4$ and apply the scaled gain $K=10^{-2}K_p$. The observer gains are $\hat L_i=S_iC_i^T$, with $S_i$ obtained from the stabilizing Riccati equation using $10^{-6}I_8$, and $L_i=10C_i^T$; the consensus gain is $\eta=20$. For each tested $\rho$, we initialize with $P_e^{(0)}=I_8$, compute a feasible $H_b$, and refine the scalar variables before alternating updates. The selected design has $\rho=0.0275$~rad and, after three iterations, gives $J_{\rm dis}=0.8278$ and $\gamma_{\min}=0.02210<\gamma_{\rm off}=0.05$.

\subsection{Results}
\label{subsec:simulation-results}

We simulate~\eqref{eq:sim-nonlinear-plant} for $400$~s under Algorithm~\ref{alg:topology-switching-protocol} and with all agents in disconnected mode. Both cases use the same bounded sinusoidal disturbance, measurement noise, and randomly chosen $x(0)\in\mathcal B$. Each observer starts from $\hat x_i(0)=0$, with $e(0)\in\Omega$.

\begin{figure}[t]
\centering
\makebox[\columnwidth][c]{\includegraphics[width=1.03\columnwidth]{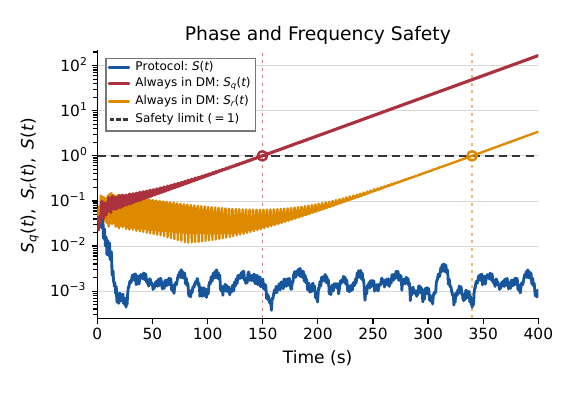}}
\caption{Normalized phase and frequency safety over $t\in[0,400]$~s. The blue curve shows $S(t)$ under the protocol; the red and orange curves show $S_q(t)$ and $S_r(t)$ with all agents in disconnected mode (DM). The horizontal dashed line marks the safety limit $1$. Open circles and vertical dotted lines mark the first phase and frequency crossings.}
\label{fig:full-state-safety}
\end{figure}

\begin{figure}[t]
\centering
\makebox[\columnwidth][c]{\includegraphics[width=1.03\columnwidth]{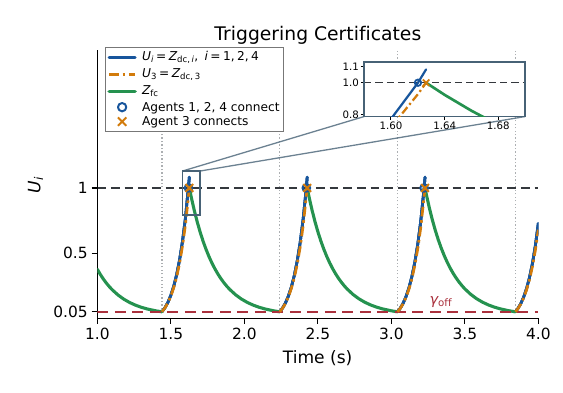}}
\caption{The local bounds $U_i$ and $Z_{\rm fc}$ over $1$--$4$~s. Under full disconnection, $U_i=Z_{{\rm dc},i}$, with the curves for agents $1$, $2$, and $4$ overlapping. Open circles and crosses mark their connection events and those of agent $3$, respectively. The horizontal dashed lines mark $1$ and $\gamma_{\rm off}=0.05$, and the vertical dotted lines mark joint disconnections. The inset shows the distinct connection times and subsequent recovery under full connectivity.}
\label{fig:triggering-certificates}
\end{figure}

Under the protocol, $S(t)\le0.05758<1$ and $V_e(t)\le0.7573<1$ for all $t\in[0,400]$~s. Figure~\ref{fig:full-state-safety} shows that, with all agents in disconnected mode, the phase and frequency limits are first exceeded at $149.73$~s and $339.95$~s, respectively. In this case, $V_e$ first exceeds $1$ at $3.11$~s. Figure~\ref{fig:triggering-certificates} shows the connection and joint disconnection events under Algorithm~\ref{alg:topology-switching-protocol}. Over $400$~s, all four agents are simultaneously in disconnected mode for $22.57\%$ of the time. Averaged over all four agents, the time spent in disconnected mode is $22.76\%$. These results demonstrate that the proposed protocol preserves state safety and reduces communication costs by allowing agents to remain in disconnected mode for part of the simulation time.

\section{Conclusion and Future Work}
\label{sec:conclusion}
\color{black}

This paper developed a decentralized event-triggered topology-switching protocol that allowed agents to disconnect from the communication network while preserving state safety and excluding Zeno behavior. We established a robust barrier function condition that guaranteed state safety as long as the estimation error remained within a prescribed threshold. Each agent used locally computable certificates to determine when to connect and disconnect from the network. We proved that the estimation error did not exceed the prescribed threshold and that the protocol was non-Zeno. The certificate construction improved scalability by reducing the complexity of topology-dependent computations from $O(2^N)$ to $O(N)$. An alternating method computed an error-weighting matrix and a quadratic barrier function that guaranteed safety under the proposed protocol. In the four-microgrid simulation, phase and frequency deviations remained within their prescribed safe operating region throughout the simulation. Compared with maintaining network connectivity at all times, the protocol reduced the average time agents remained connected by $22.76\%$, with all four agents simultaneously disconnected for $22.57\%$ of the simulation. Future work will derive safety conditions for nonlinear multi-agent systems and adapt the protocol to handle delayed readiness bits piggybacked on routing messages and asynchronous topology switching, in which agents in the same connected component may disconnect at different times.
\appendices
\section{Proof of Lemma~\ref{lem:error-comparison-bound}}
\label{app:error-comparison-bound}

\begin{proof}
Recall~\eqref{eq:dynamics-error}. Differentiating $V_e(e)=e^\top\bar P e$ gives
\begin{equation}
\begin{aligned}
\dot V_e
&=\dot e^\top\bar P e+e^\top\bar P\dot e\\
&=\underbrace{
e^\top\bigl[A_e(\mathcal L_0)^\top\bar P+\bar P A_e(\mathcal L_0)\bigr]e
}_{\text{Term I}}\\
&\quad+\underbrace{2e^\top\bar P\mathcal Iw}_{\text{Term II}}
+\underbrace{\bigl(-2e^\top\bar P J(\mathcal L_0)v\bigr)}_{\text{Term III}}.
\end{aligned}
\label{eq:Ve-derivative-decomposition}
\end{equation}
The three terms in~\eqref{eq:Ve-derivative-decomposition} relate to the error dynamics, process disturbance, and measurement noise, respectively.

For Term~I, substituting $e=\bar P^{-1/2}(\bar P^{1/2}e)$ and using~\eqref{eq:mu-definition} gives $\text{Term I}\leq\mu(\mathcal L_0)\|\bar P^{1/2}e\|_2^2=\mu(\mathcal L_0)V_e(e)$. Equivalently, $A_e(\mathcal L_0)^\top\bar P+\bar P A_e(\mathcal L_0)\preceq\mu(\mathcal L_0)\bar P$.

For Term~II, inserting $Q_w^{-1/2}Q_w^{1/2}=I$, applying the Cauchy--Schwarz inequality, and using $w^\top Q_w w\leq1$ gives
\[
\begin{aligned}
\text{Term II}
&\leq 2r_e\|\bar P^{1/2}\mathcal I Q_w^{-1/2}\|_2\|Q_w^{1/2}w\|_2\\
&\leq 2r_e\|\bar P^{1/2}\mathcal I Q_w^{-1/2}\|_2.
\end{aligned}
\]
Similarly, using $R_v^{-1/2}R_v^{1/2}=I$ and $v^\top R_vv\leq1$ gives
\[
\begin{aligned}
\text{Term III}
&\leq 2r_e\|\bar P^{1/2}J(\mathcal L_0)R_v^{-1/2}\|_2\|R_v^{1/2}v\|_2\\
&\leq 2r_e\|\bar P^{1/2}J(\mathcal L_0)R_v^{-1/2}\|_2.
\end{aligned}
\]
Combining these bounds yields $\dot V_e\leq\mu(\mathcal L_0)V_e+2c(\mathcal L_0)r_e$. Integrating this bound and using continuity gives the same bound for $D^+V_e$.

When $r_e(t)>0$, the relation $V_e=r_e^2$ gives
\[D^+r_e(t)=\frac{D^+V_e(e(t))}{2r_e(t)}
\leq\frac12\mu(\mathcal L_0)r_e(t)+c(\mathcal L_0).\]

When $r_e(t)=0$, i.e., $e(t)=0$, division by $2r_e(t)$ is not defined. For $h>0$ with $t+h<t_4$, the disturbance bounds in~\eqref{eq:disturbance-set} and~\eqref{eq:noise-set} give 
\[
\begin{aligned}
&\bigl\|\bar P^{1/2}
  [\mathcal Iw(s)-J(\mathcal L_0)v(s)]\bigr\|_2\\
&\quad\leq
\|\bar P^{1/2}\mathcal I Q_w^{-1/2}\|_2
\|Q_w^{1/2}w(s)\|_2\\
&\qquad+
\|\bar P^{1/2}J(\mathcal L_0)R_v^{-1/2}\|_2
\|R_v^{1/2}v(s)\|_2\\
&\quad\leq
\|\bar P^{1/2}\mathcal I Q_w^{-1/2}\|_2+
\|\bar P^{1/2}J(\mathcal L_0)R_v^{-1/2}\|_2\\
&\quad=c(\mathcal L_0)
\end{aligned}
\]
for $s\in[t,t+h]$. Since $e(t)=0$, integrating~\eqref{eq:dynamics-error} and applying the triangle inequality give
\[
\begin{aligned}
r_e(t+h)
&=\|\bar P^{1/2}e(t+h)\|_2\\
&=\left\|\int_t^{t+h}\bar P^{1/2}\dot e(s)\,ds\right\|_2\\
&\leq\int_t^{t+h}\|\bar P^{1/2}\dot e(s)\|_2\,ds\\
&\leq\int_t^{t+h}\|\bar P^{1/2}A_e(\mathcal L_0)e(s)\|_2\,ds\\
&\quad+\int_t^{t+h}
\|\bar P^{1/2}[\mathcal Iw(s)-J(\mathcal L_0)v(s)]\|_2\,ds\\
&\leq\int_t^{t+h}\|\bar P^{1/2}A_e(\mathcal L_0)e(s)\|_2\,ds
+c(\mathcal L_0)h.
\end{aligned}
\]
Dividing by $h>0$ gives
\[
\begin{aligned}
\frac{r_e(t+h)}{h}
&\leq c(\mathcal L_0)
+\frac1h\int_t^{t+h}\|\bar P^{1/2}A_e(\mathcal L_0)e(s)\|_2\,ds\\
&\leq c(\mathcal L_0)
+\sup_{s\in[t,t+h]}\|\bar P^{1/2}A_e(\mathcal L_0)e(s)\|_2.
\end{aligned}
\]
Since $e$ is continuous and $e(t)=0$, the supremum tends to $0$ as $h\to0^+$. Hence,
\[D^+r_e(t)=\limsup_{h\to0^+}\frac{r_e(t+h)-r_e(t)}{h}
\leq c(\mathcal L_0).\] So,~\eqref{eq:error-radius-differential-bound} holds for all $t\in[t_3,t_4)$.
\end{proof}

\section{Proof of Lemma~\ref{lem:local-comparison-coefficients}}
\label{app:local-comparison-coefficients}

\begin{proof}
Fix $i\in\mathcal N$ and $t\in[t_7,t_8)$, and write $\mathcal L(t)=\mathcal L(t_7)=\mathcal L_2$. By~\eqref{Fij} and~\eqref{eq:normalized-F-matrix}, $M_{jk}(\mathcal L_2)=M_{jk}(0)$ for $j\neq k$, and
\[
\begin{aligned}
&M_{kk}(\mathcal L_2)-M_{kk}(0)\\
&\quad=-P_e^{-1/2}\Bigl[
C_k^\top(\bar L_k(\ell_{kk})-\hat L_k)^\top P_e\\
&\qquad\qquad+P_e(\bar L_k(\ell_{kk})-\hat L_k)C_k
\Bigr]P_e^{-1/2}\\
&\quad=
\begin{cases}
0, & \ell_{kk}=0,\\
\Delta M_k, & \ell_{kk}>0.
\end{cases}
\end{aligned}
\]

If $\ell_{ii}=0$, the $i$th diagonal block of $M(\mathcal L_2)-M(0)$ is zero. Using its block-diagonal structure and $-2\eta(\mathcal L_2\otimes I_n)\preceq0$ gives
\[
\begin{aligned}
\mu(\mathcal L_2)
&=\lambda_{\max}\!\bigl(
M(\mathcal L_2)-2\eta(\mathcal L_2\otimes I_n)
\bigr)\\
&\leq\lambda_{\max}(M(\mathcal L_2))\\
&\leq\lambda_{\max}(M(0))
+\lambda_{\max}\!\bigl(M(\mathcal L_2)-M(0)\bigr)\\
&\leq\lambda_{\max}(M(0))
+\max\!\left\{0,\max_{k\in\mathcal N\setminus\{i\}}
\lambda_{\max}(\Delta M_k)\right\}\\
&=\bar\mu_i^{\mathrm{dis}}.
\end{aligned}
\]

If $\ell_{ii}>0$, each $k\in\mathcal R_i(t)\setminus\{i\}$ has a neighbor along a communication path from $i$, so $\ell_{kk}>0$ for every $k\in\mathcal R_i(t)$. The block identities above therefore give
\[
\begin{aligned}
&\lambda_{\max}\!\Bigl(
M(\mathcal L_2)-M(0)
-\sum_{k\in\mathcal R_i(t)}\Pi_k^\top\Delta M_k\Pi_k
\Bigr)\\
&\quad=\lambda_{\max}\!\Bigl(
\sum_{\substack{k\in\mathcal N\setminus\mathcal R_i(t)\\
\ell_{kk}>0}}
\Pi_k^\top\Delta M_k\Pi_k
\Bigr)\\
&\quad\leq
\max\!\left\{0,\max_{k\in\mathcal N\setminus\mathcal R_i(t)}
\lambda_{\max}(\Delta M_k)\right\}.
\end{aligned}
\]
Here, $\mathcal N\setminus\mathcal R_i(t)\neq\varnothing$ because $\mathcal R_i(t)\subseteq\Psi(t)\neq\mathcal N$. Since no agent in $\mathcal R_i(t)$ has a neighbor in $\mathcal N\setminus\mathcal R_i(t)$, $\mathcal L_2-\mathcal L_i^{\mathrm{known}}(t)\succeq0$ is the Laplacian of the remaining active communication edges.
Since $\eta\geq0$, we have $-2\eta((\mathcal L_2-\mathcal L_i^{\mathrm{known}}(t))\otimes I_n)\preceq0$. Thus,
\[
\begin{aligned}
\mu(\mathcal L_2)
&=\lambda_{\max}\!\bigl(
M(\mathcal L_2)-2\eta(\mathcal L_2\otimes I_n)
\bigr)\\
&=\lambda_{\max}\!\Bigl(
M(\mathcal L_2)-2\eta(\mathcal L_i^{\mathrm{known}}(t)\otimes I_n)\\
&\qquad-2\eta\bigl(
(\mathcal L_2-\mathcal L_i^{\mathrm{known}}(t))\otimes I_n
\bigr)\Bigr)\\
&\leq\lambda_{\max}\!\bigl(
M(\mathcal L_2)-2\eta(\mathcal L_i^{\mathrm{known}}(t)\otimes I_n)
\bigr)\\
&\leq\lambda_{\max}\!\Bigl(
M(0)-2\eta(\mathcal L_i^{\mathrm{known}}(t)\otimes I_n)\\
&\qquad+\sum_{k\in\mathcal R_i(t)}
\Pi_k^\top\Delta M_k\Pi_k
\Bigr)\\
&\quad+\max\!\left\{0,
\max_{k\in\mathcal N\setminus\mathcal R_i(t)}
\lambda_{\max}(\Delta M_k)\right\}\\
&=\bar\mu_i^{\mathrm{con}}(t).
\end{aligned}
\]

By~\eqref{eq:c-definition}, the block-diagonal norm identity, and~\eqref{mode-dependent gain},
\[
\begin{aligned}
c(\mathcal L_2)
&=\|\bar P^{1/2}\mathcal I Q_w^{-1/2}\|_2\\
&\quad+\max_{k\in\mathcal N}
\|P_e^{1/2}\bar L_k(\ell_{kk})R_{v,k}^{-1/2}\|_2\\
&\leq
\begin{cases}
\bar c_i^{\mathrm{dis}}, & \ell_{ii}=0,\\
\bar c_i^{\mathrm{con}}(t), & \ell_{ii}>0.
\end{cases}
\end{aligned}
\]
In the first case, $\bar L_i=\hat L_i$ and $\bar L_k\in\{\hat L_k,NL_k\}$ for $k\neq i$. In the second case, $\bar L_k=NL_k$ for $k\in\mathcal R_i(t)$ and $\bar L_k\in\{\hat L_k,NL_k\}$ otherwise. These are precisely the gain choices included in the definitions of $\bar c_i^{\mathrm{dis}}$ and $\bar c_i^{\mathrm{con}}(t)$.

Finally, the coefficients are constant on $[t_7,t_8)$. Since $r_e(t)\geq0$, the bounds on $\mu(\mathcal L_2)$ and $c(\mathcal L_2)$ preserve the radius inequality. For either coefficient pair, the scalar comparison principle applied to the radius given by~\eqref{eq:comparison-radius-solution}, with $\bar r_{e,i}(t_7)\geq r_e(t_7)$, yields $0\leq r_e(t)\leq\bar r_{e,i}(t)$. Therefore, $V_e(e(t))=r_e^2(t)\leq\bar r_{e,i}^{\,2}(t)=U_i(t)$ for all $t\in[t_7,t_8)$.
\end{proof}

\bibliographystyle{IEEEtran}
\bibliography{references}

@ARTICLE{johansson2000,
  author={Johansson, K.H.},
  journal={IEEE Transactions on Control Systems Technology}, 
  title={The quadruple-tank process: a multivariable laboratory process with an adjustable zero}, 
  year={2000},
  volume={8},
  number={3},
  pages={456-465},
  doi={10.1109/87.845876},
  ISSN={1558-0865},
  month={May},}

@ARTICLE{10360262,
  author={Griffioen, Paul and Romagnoli, Raffaele and Krogh, Bruce H. and Sinopoli, Bruno},
  journal={IEEE Transactions on Control of Network Systems}, 
  title={Reducing Attack Opportunities Through Decentralized Event-Triggered Control}, 
  year={2024},
  volume={11},
  number={3},
  pages={1576-1586},
  doi={10.1109/TCNS.2023.3343151}}

@inproceedings{dai2023convex,
  title={Convex synthesis and verification of control-Lyapunov and barrier functions with input constraints},
  author={Dai, Hongkai and Permenter, Frank},
  booktitle={2023 American Control Conference (ACC)},
  pages={4116--4123},
  year={2023},
  organization={IEEE}
}

@INPROCEEDINGS{6425820,
  author={Heemels, W.P.M.H. and Johansson, K.H. and Tabuada, P.},
  booktitle={2012 IEEE 51st IEEE Conference on Decision and Control (CDC)}, 
  title={An introduction to event-triggered and self-triggered control}, 
  year={2012},
  volume={},
  number={},
  pages={3270-3285},
  doi={10.1109/CDC.2012.6425820}}

@book{bullo2009distributed,
  title={Distributed control of robotic networks: a mathematical approach to motion coordination algorithms},
  author={Bullo, Francesco and Cort{\'e}s, Jorge and Martinez, Sonia},
  year={2009},
  publisher={Princeton University Press}
}

@article{deng2022distributed,
  title={Distributed adaptive tracking control for high-order nonlinear multiagent systems over event-triggered communication},
  author={Deng, Chao and Wen, Changyun and Wang, Wei and Li, Xinyao and Yue, Dong},
  journal={IEEE Transactions on Automatic Control},
  volume={68},
  number={2},
  pages={1176--1183},
  year={2022},
  publisher={IEEE}
}

@article{nowzari2019event,
  title={Event-triggered communication and control of networked systems for multi-agent consensus},
  author={Nowzari, Cameron and Garcia, Eloy and Cort{\'e}s, Jorge},
  journal={Automatica},
  volume={105},
  pages={1--27},
  year={2019},
  publisher={Elsevier}
}

@article{an2023event,
  title={Event-triggered reference governors for collisions-free leader--following coordination under unreliable communication topologies},
  author={An, Liwei and Yang, Guang-Hong and Deng, Chao and Wen, Changyun},
  journal={IEEE Transactions on Automatic Control},
  volume={69},
  number={4},
  pages={2116--2130},
  year={2023},
  publisher={IEEE}
}

@article{ge2021dynamic,
  title={Dynamic event-triggered control and estimation: A survey},
  author={Ge, Xiaohua and Han, Qing-Long and Zhang, Xian-Ming and Ding, Derui},
  journal={International Journal of Automation and Computing},
  volume={18},
  number={6},
  pages={857--886},
  year={2021},
  publisher={Springer}
}

@article{rajaraman2002topology,
  title={Topology control and routing in ad hoc networks: A survey},
  author={Rajaraman, Rajmohan},
  journal={ACM SIGACT News},
  volume={33},
  number={2},
  pages={60--73},
  year={2002},
  publisher={ACM New York, NY, USA}
}

@inproceedings{sabouni2023optimal,
  title={Optimal control of connected automated vehicles with event-triggered control barrier functions: a test bed for safe optimal merging},
  author={Sabouni, Ehsan and Ahmad, HM Sabbir and Xiao, Wei and Cassandras, Christos G and Li, Wenchao},
  booktitle={2023 IEEE Conference on Control Technology and Applications (CCTA)},
  pages={321--326},
  year={2023},
  organization={IEEE}
}

@article{luo2020multi,
  title={Multi-robot collision avoidance under uncertainty with probabilistic safety barrier certificates},
  author={Luo, Wenhao and Sun, Wen and Kapoor, Ashish},
  journal={Advances in Neural Information Processing Systems},
  volume={33},
  pages={372--383},
  year={2020}
}

@inproceedings{Wang_2016,
   title={Safety barrier certificates for heterogeneous multi-robot systems},
   DOI={10.1109/acc.2016.7526486},
   booktitle={2016 American Control Conference (ACC)},
   publisher={IEEE},
   author={Wang, Li and Ames, Aaron and Egerstedt, Magnus},
   year={2016},
   month=July, pages={5213–5218} }

@article{peng2013event,
  title={Event-triggered communication and {$H_\infty$} control co-design for networked control systems},
  author={Peng, Chen and Yang, Tai Cheng},
  journal={Automatica},
  volume={49},
  number={5},
  pages={1326--1332},
  year={2013},
  publisher={Elsevier}
}

@INPROCEEDINGS{8796030,
  author={Ames, Aaron D. and Coogan, Samuel and Egerstedt, Magnus and Notomista, Gennaro and Sreenath, Koushil and Tabuada, Paulo},
  booktitle={2019 18th European Control Conference (ECC)}, 
  title={Control Barrier Functions: Theory and Applications}, 
  year={2019},
  volume={},
  number={},
  pages={3420-3431},
  doi={10.23919/ECC.2019.8796030},
  ISSN={},
  month={June},}

@ARTICLE{9507261,
  author={Gholami, Amin and Sun, Xu Andy},
  journal={IEEE Transactions on Control of Network Systems}, 
  title={Stability of Multimicrogrids: New Certificates, Distributed Control, and Braess’s Paradox}, 
  year={2022},
  volume={9},
  number={1},
  pages={308-319},
  doi={10.1109/TCNS.2021.3102006}}

@inproceedings{wei2026verifiablysafedecentralizedcontrol,
  author    = {Amber He Wei and Shiyu Cheng and Andrew Clark and Bruno Sinopoli},
  title     = {Verifiably Safe Decentralized Control via Event-Triggered Topology Switching},
  booktitle = {Proceedings of the 62nd Allerton Conference on Communication, Control, and Computing},
  year      = {2026},
  note      = {To appear},
  url       = {https://arxiv.org/abs/2611.23415}
}

@ARTICLE{ames2017control,
  author={Ames, Aaron D. and Xu, Xiangru and Grizzle, Jessy W. and Tabuada, Paulo},
  journal={IEEE Transactions on Automatic Control}, 
  title={Control Barrier Function Based Quadratic Programs for Safety Critical Systems}, 
  year={2017},
  volume={62},
  number={8},
  pages={3861-3876},
  doi={10.1109/TAC.2016.2638961}}

@INPROCEEDINGS{wu2016safety,
  author={Wu, Guofan and Sreenath, Koushil},
  booktitle={2016 American Control Conference (ACC)}, 
  title={Safety-critical control of a planar quadrotor}, 
  year={2016},
  volume={},
  number={},
  pages={2252-2258},
  doi={10.1109/ACC.2016.7525253}}

@INPROCEEDINGS{9636584,
  author={Cosner, Ryan K. and Singletary, Andrew W. and Taylor, Andrew J. and Molnar, Tamas G. and Bouman, Katherine L. and Ames, Aaron D.},
  booktitle={2021 IEEE/RSJ International Conference on Intelligent Robots and Systems (IROS)}, 
  title={Measurement-Robust Control Barrier Functions: Certainty in Safety with Uncertainty in State}, 
  year={2021},
  volume={},
  number={},
  pages={6286-6291},
  doi={10.1109/IROS51168.2021.9636584}}

@INPROCEEDINGS{9867262,
  author={Wang, Yujie and Xu, Xiangru},
  booktitle={2022 American Control Conference (ACC)}, 
  title={Observer-based Control Barrier Functions for Safety Critical Systems}, 
  year={2022},
  volume={},
  number={},
  pages={709-714},
  doi={10.23919/ACC53348.2022.9867262}}
\end{document}